\documentclass[pra,aps,twocolumn,superscriptaddress,longbibliography]{revtex4-1}
\usepackage[colorlinks=true, citecolor=blue, urlcolor=magenta ]{hyperref}
\usepackage{bm}
\usepackage{amsmath,amsfonts}
\usepackage{amsthm}
\usepackage{hyperref}
\usepackage{braket}
\theoremstyle{plain}
\usepackage{color}
\usepackage{amssymb}
\usepackage{amsthm}
\usepackage{float}
\usepackage{mathtools}
\usepackage{esvect}
\usepackage{wrapfig}
\usepackage{amsthm}
\usepackage{verbatim}
\usepackage{bbm}
\usepackage[normalem]{ulem}
\usepackage{enumitem}
\usepackage{fmtcount}
\usepackage{booktabs}
\usepackage{csquotes}
\usepackage{epsfig}
\usepackage{tabularx}
\usepackage{graphicx}
\usepackage{braket}
\usepackage{latexsym}
\usepackage{bm}
\usepackage{graphics,epstopdf}
\usepackage{enumitem}
\usepackage{fmtcount}
\usepackage{booktabs}
\usepackage{csquotes}
\usepackage{epsfig}
\theoremstyle{plain}
\usepackage{float}
\usepackage{graphicx}  
\usepackage{dcolumn}          
\usepackage{amssymb}
\usepackage{appendix}
\usepackage{physics}   
\usepackage{mathtools}
\usepackage{esvect}
\usepackage{wrapfig}
\usepackage{amsthm}
\usepackage{verbatim}
\usepackage{bbm}

\usepackage[mathscr]{euscript}
\def\Tr{\operatorname{Tr}}

\begin{document}

\title{Resourcefulness vs indivisibility in quantum channels}

\author{Priya Ghosh}
\affiliation{Harish-Chandra Research Institute,  A CI of Homi Bhabha National Institute, Chhatnag Road, Jhunsi, Prayagraj  211019, India}
\author{Soumyajit Pal}
\affiliation{Harish-Chandra Research Institute,  A CI of Homi Bhabha National Institute, Chhatnag Road, Jhunsi, Prayagraj  211019, India}
\affiliation{Department of Physical Sciences, IISER Berhampur, Berhampur  760010, India}
\author{Ujjwal Sen}
\affiliation{Harish-Chandra Research Institute,  A CI of Homi Bhabha National Institute, Chhatnag Road, Jhunsi, Prayagraj  211019, India}

\begin{abstract}
Open quantum dynamics 
can be categorized in several ways, including according to their divisibility.  For any quantum channel acting for any finite time period, we propose measures of P-indivisibility and CP-indivisibility, where P and CP stand for positivity and complete positivity respectively. 
Subsequently, we also propose two quantities 
to measure the resourcefulness - with respect to an arbitrary  quantum resource - of any quantum channel within any finite time interval.
Moreover, we find a bridge between these two classes of metrics, viz. those quantifying divisibilities of quantum channels and those gauging their resourcefulness, by identifying two separate relations between elements of one class with those of the other. 
Lastly, we verify the two relations using quantum non-Markovianity as a channel resource.
\end{abstract}

\maketitle

\section{Introduction}
\label{sec1}

In the real world, noise is an unavoidable presence, often rendering closed system analyses impractical. Consequently, it becomes imperative to study systems in the context of their environment, giving rise to what is known as open system dynamics~\cite{Rivas-open,RHP-review,BLPV-review,VA-review}. Open quantum dynamics can be classified in a variety of ways, including 
according to either P-indivisibility or CP-indivisibility, where CP stands for complete positivity and P stands for positivity. Furthermore, measures of CP-indivisibility~\cite{RHP-1} and P-indivisibility~\cite{BLP-1} for quantum channels have been proposed to quantify these properties. 
On the other hand, a quantum map is termed as Markovian~\cite{petru-book} if, after its action on the system's state, the output state becomes uncorrelated with the environment, and the environment remains unchanged. Otherwise, the quantum map is labeled as non-Markovian. However, the precise mathematical definition of Markovian maps remains elusive. Sometimes, CP-divisible maps or P-divisible maps are referred to as Markovian maps. Moreover, several measures for quantifying non-Markovianity~\cite{marko-measure-1,BLP-1,RHP-1,marko-measure-4,marko-measure-5,marko-measure-6,marko-measure-7,marko-measure-8,marko-measure-9,marko-measure-10,marko-measure-11,marko-measure-12,marko-measure-13,marko-measure-14,marko-measure-15} in quantum dynamics have been developed.

Quantum information tasks are often associated with a specific quantum resource, which is necessary either for achieving success in the task or for attaining a quantum advantage over the corresponding classical methods.
Various quantum properties, such as entanglement~\cite{ent-re-1,ent-re-2,ent-re-3,ent-re-4,ent-re-5}, purity~\cite{thermo-re-2,purity-re-2}, quantum coherence~\cite{coherence-re-1,coherence-re-2,coherence-re-3}, asymmetry~\cite{asymmetry-re-1,asymmetry-re-2,asymmetry-re-3}, magic~\cite{magic-re-1,magic-re-2,magic-re-3}, athermality~\cite{thermo-re-1,thermo-re-2,thermo-re-3,thermo-re-4,thermo-re-5,thermo-re-6,thermo-re-7,thermo-re-8}, and so on serve as quantum resources across a spectrum of quantum information tasks, and 
%. In the field of quantum physics, quantum 
resource theories~\cite{re-1,re-book-2} have been developed for each quantum resource. These theories aid in the classification of states or channels and also help to find quantifiers to measure the resourcefulness of any quantum resource present in states or channels. Additionally, quantum resource theories facilitate the identification of possible inter-conversions among quantum states and channels, and even establish connections between different quantum resource theories.

Most, though not all, resource theories focus on quantifying the resourcefulness - with respect to a particular resource -  
of quantum states. However, in our study, we concentrate on quantifying the resourcefulness of 
quantum channels. Corresponding to an arbitrary open system dynamics, we introduce two metrics designed to quantify the resourcefulness - with respect to an arbitrary quantum resource - 
of any quantum dynamics within any finite time interval.
Furthermore, since  quantum maps within a finite time interval can be categorized by its indivisibility in the open system dynamics, we propose measures of CP-indivisibility and P-indivisibility for any quantum channel within any finite time interval. To our knowledge, the relationship between measures of resourcefulness of any quantum resource for any quantum channel and its measures of indivisibility remains unexplored in the existing literature. In our work, we address this question, and find
two relationships: one between the measure of resourcefulness of any quantum resource and the measure of P-indivisibility for any quantum channel within any finite time interval, and another between the measure of resourcefulness of any quantum resource and the measure of CP-indivisibility for any quantum channel within any finite time interval.
Lastly, we present an example of a quantum channel for which we measure the resourcefulness of ``quantum non-Markovianity" to validate the relationships uncovered.

The remainder of the paper is structured as follows. In Section~\ref{sec2}, we provide a brief overview of P-indivisible and CP-indivisible quantum maps and propose metrics for quantifying each type of indivisibility  for quantum channels within finite time intervals. We then introduce two metrics for quantifying the resourcefulness of any quantum resource for quantum channels within finite time intervals. 
In Section~\ref{sec3}, we establish connections between the metrics for quantifying the resourcefulness of any quantum property for quantum channels and the metrics for quantifying their indivisibility within finite time intervals in the framework of open system dynamics.
Subsequently, in Section~\ref{sec4}, we illustrate our findings by presenting an example of a quantum channel using  ``quantum non-Markovianity" as a quantum resource.
Finally, we present the concluding remarks in Section~\ref{sec5}.

\section{Preliminaries}
\label{sec2}
Let $\mathcal{H}$ represent the Hilbert space of quantum states of a system, and we will use its suffix to represent the corresponding system. $\mathcal{D}(\mathcal{H})$ represents a set of all quantum states on the Hilbert space $\mathcal{H}$.
A quantum map is a superoperator that acts upon a quantum state on the Hilbert space $\mathcal{H}$ and changes it into a quantum state on the same Hilbert space. A quantum map operating on a quantum state essentially depicts the density matrix's or the quantum state's dynamic evolution. A quantum map can also be called a ``quantum channel," ``quantum process," or ``quantum operation". Here, we represent a quantum map using the notation $``\Lambda"$.
Let us assume that $\rho(t)$ represents a quantum state of a system at time $t$, and $\Lambda (t_2,t_1)$ represents a quantum map that evolves any quantum state of a system from time $t_1$ to time $t_2$ with $t_2 \geq t_1 \geq 0$. The transformation of a
system's state from $\rho(t_1)$ to $\rho(t_2) $ by a quantum map $\Lambda (t_2,t_1)$ can be expressed as:
\begin{align}
\rho (t_2) \coloneqq \Lambda (t_2,t_1) \rho (t_1),
\end{align}
where $t_2 \geq t_1 \geq 0$.

In the rest of the paper, we will focus on the open system dynamics of any system. By ``open system dynamics," we mean that the system, together with the environment, forms a closed composite system, and the composite system goes through unitary evolution. $\mathcal{H}_S$ and $\mathcal{H}_E$ will represent the Hilbert space of our system and environment, respectively, and $\mathcal{H}_S \otimes \mathcal{H}_E$ will represent the Hilbert space of a composite system consisting of our system and environment.
Hence, in open system dynamics, we cannot represent the quantum map acting on the quantum state of the system by the unitary evolution of the state. In open system dynamics, the evolved state of the system after the action of a quantum map on the quantum state will be the state generated by taking the partial trace over the environmental part of the state of the composite system, followed by the unitary evolution of the system-environment initial state of the composite system.
Mathematically, the quantum state after the action of the quantum map, say $\Lambda(t,0)$, on a quantum state $\rho(0)$ of a system, will be
\begin{align}
\Lambda(t,0) (\rho(0)) = \Tr_E[U \rho_{SE}(0) U^\dagger]
\end{align}
where $\Tr_E$ denotes the partial trace over the environmental part from the state of the composite system and $U$ denotes the unitary evolution of the composite system. $\rho_{SE}(0)$ denotes the system-environment initial state of the composite system, and $\rho(0) = \Tr_E [\rho_{SE}(0)] $.

In open system dynamics, a quantum map turns into a completely positive (CP) map in a particular scenario, that is when the system is initially uncorrelated with the environment. Since any CP map can be decomposed into Kraus operators, any quantum map can be written in terms of Kraus operators when the system-environment initial state of a composite system is in product form. It was shown that the Kraus operator decomposition of any quantum CP map is not unique.

We can classify the set of all quantum dynamics into P-divisible maps and P-indivisible maps.
Any map $\Lambda$ will be called P-divisible if we can decompose the map as follows:
\begin{equation}
\label{eq-div-1}
\Lambda(t+\tau, 0) = \Lambda(t+\tau, t) \cdot \Lambda(t, 0),
\end{equation}
for all times $t$ , $\tau \geq 0$ and $\Lambda ( t + \tau , t )$ is positive and trace preserving for all intermediate times. Otherwise, the map will be called P-indivisible one.
In Ref.~\cite{BLP-1,P-div-inf-backflow}, it was shown that the trace distance between $\Lambda ( \rho_1 )$ and $\Lambda ( \rho_2 )$
%and $\Lambda(\rho_2)$ 
will decrease monotonically with time if and only if the map $\Lambda$ is P-divisible map, whatever two initial quantum states of the system, $\rho_1$ and $\rho_2$, on $\mathcal{H}_S$ are considered.
The trace distance between two quantum states, $\rho_1$ and $\rho_2$, is defined as 
\begin{align*} 
D [\rho_1 || \rho_2] \coloneqq \frac{1}{2} ||\rho_1 -\rho_2||_1, \end{align*} 
where $||A||_1$ denotes the trace norm of matrix $A$ or equal to  $\sqrt{A^\dagger A}$ for matrix $A$. Based on this, we can define a quantifier of P-indivisibility for any quantum map $\Lambda$ in between the time interval $\lbrack t, t + \tau \rbrack$ where $t, \tau \geq 0$, and we will refer to this measure as ``$\textnormal{P-I} (\Lambda, t, \tau)$". $\textnormal{P-I} (\Lambda, t, \tau)$ will be a function of $\Lambda$, $t$, and $\tau$, and it can be mathematically expressed as follows: 
\begin{align} \label{eqn-def-P-I} & \textnormal{P-I} (\Lambda, t, \tau) \coloneqq \nonumber \\ &\max \Big\{ \max_{\rho_1, \rho_2 \in \mathcal{D}(\mathcal{H_S})} \lbrace D [\Lambda (t+\tau,0) \rho_1 || \Lambda (t+\tau,0) \rho_2] - \nonumber \\ & \phantom{ami bhat khabo, ami ba}D [\Lambda (t,0) \rho_1 || \Lambda (t,0) \rho_2] \rbrace, 0 \Big\}. \end{align} 
Here, $\rho_1$ and $\rho_2$ represent any two arbitrary initial states of the system on $\mathcal{H}_S$.

Similarly, one can categorize the set of quantum dynamics into two parts: CP-divisible and CP-indivisible maps. Any map $\Lambda$ will be called CP-divisible if we can write the map as Eq.~\eqref{eq-div-1} for all times and the map at any intermediate time is completely positive trace preserving; otherwise, it will be called a CP-indivisible map. 
It can be easily seen that the set of CP-divisible maps is a subset of the set of P-divisible maps. Hence, a CP-divisible map is also a P-divisible map, but the reverse is not always true.
From the definition of CP-divisibility, it can be easily proved that a map will be a CP-divisible one if and only if $|| (\Lambda(t+\tau,t) \otimes \mathbbm{1}) \ket{\phi^{+}} \bra{\phi^{+}}||_1$ is equal to $1$, where $||A||_1$ denotes the trace norm of matrix $A$~\cite{RHP-1} and $\ket{\phi^{+}} \bra{\phi^{+}}$ denotes the maximally entangled state between the system and environment. The symbol $\mathbbm{1}$ signifies the identity operator on the Hilbert space $\mathcal{H}_E$. From this, we can also define a measure of CP-indivisibility for a quantum map $\Lambda$ within the time interval $\lbrack t, t+\tau \rbrack$ as follows: \begin{align} \label{eqn-def-CP-I} \textnormal{CP-I} (\Lambda, t, \tau) \coloneqq || (\Lambda(t+\tau,t) \otimes \mathbbm{1}) \ket{\phi^{+}} \bra{\phi^{+}}||_1 - 1, \end{align} where $t, \tau \geq 0$.

Let's now move on to discuss the framework of quantum resource theories~\cite{re-1,re-book-2}.
When a quantum property is necessary for success in a quantum information task or to gain a quantum edge over classical ones in the task, it is referred to as a quantum resource.
A quantum resource theory has been constructed for each quantum resource, which helps to find the quantifiers of resourcefulness of the quantum resource for any given state or channel as well as feasible inter-conversions between states and so forth.
Any quantum resource theory is generally characterized by free states and free operations.
In the resource theory of any quantum resource, free states are those which have no resource. Free operations are those maps which cannot create resource from a free state, that means they transform a free state into a free state. For example, in the resource theory of entanglement, separable states act as free states, and local operations with classical communications (LOCC) act as free operations.
In general, one looks for a measure of the resourcefulness of a quantum resource, say $\mathcal{K}$, for any quantum state. 
By ``for any quantum state", we mean that how much the quantum resource is present in a quantum state and in that case the measure of resourcefulness of the quantum resource, $\mathcal{K}$, is zero for free states with respect to the resource theory of quantum resource $\mathcal{K}$.
In our paper, we will look for the measures of resourcefulness of a quantum resource for any quantum dynamics within any finite time interval since the quantum channels are generally time-dependent in open system dynamics. 
By ``for any quantum dynamics", we mean that how much maximum quantum resource can be created by the quantum map acting on quantum states where maximization is taken over all quantum states. 
For example, the measure of resourcefulness of the quantum resource $\mathcal{K}$ for free operations in the resource theory of the quantum resource $\mathcal{K}$ is zero.

Let's denote a quantum map by $\Lambda_{\mathcal{K}}$, which acts as the free operation in the resource theory of quantum resource $\mathcal{K}$ and the set of all such quantum maps, $\Lambda_{\mathcal{K}}$ by $\mathcal{D}(\Lambda)$. Now, we define a quantifier or measure of resourcefulness of the quantum resource $\mathcal{K}$ based on distance measure from the perspective of quantum maps as follows: A quantifier or measure of resourcefulness of the quantum resource $\mathcal{K}$ for any quantum map $\Lambda$ in between the time interval $[t, t + \tau]$ will be \begin{align} & \textnormal{NM}_1 (\Lambda, t, \tau) \coloneqq \nonumber \\ & \max_{\rho \in \mathcal{D}(\mathcal{H}_S)} \min_{\Lambda_{\mathcal{K}} \in \mathcal{D}(\Lambda)} T \left[ \Lambda (t+\tau,t) \Lambda (t,0) \rho || \Lambda_{\mathcal{K}} (t+\tau,t) \Lambda (t,0) \rho \right], \end{align} where $t$, $\tau \geq 0$. Here, $\rho$ represents any state of the system belonging to the set $\mathcal{D}(\mathcal{H}_S)$ and $T\left[ \rho_1 || \rho_2 \right] $ quantifies the distance between two quantum states $\rho_1$ and $\rho_2$, for example, trace distance.

\begin{figure}[h!]
\includegraphics[scale=0.45]{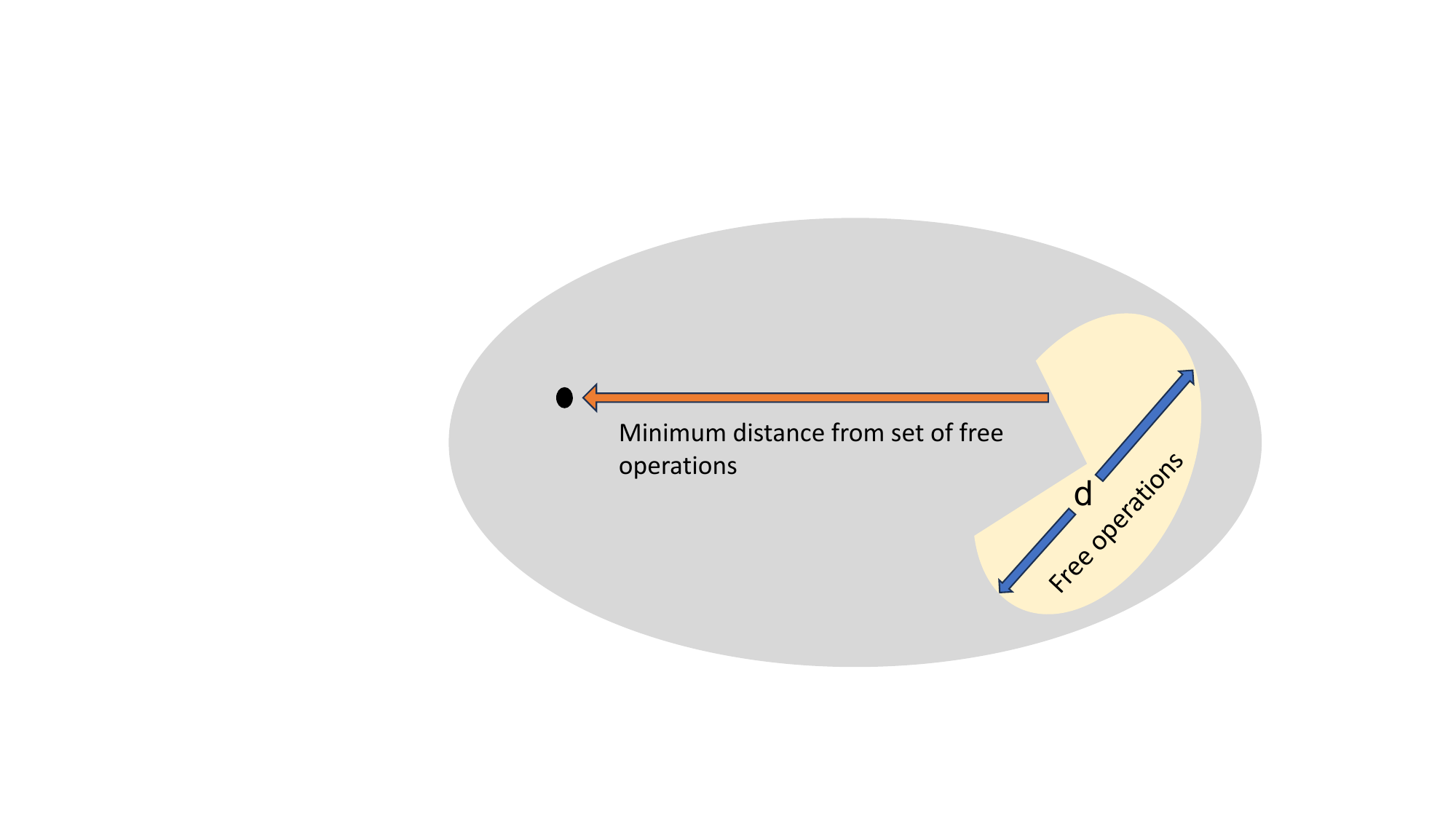}
\caption{(Color online.) \textbf{Schematic representation of the measure of resourcefulness of a quantum resource for quantum dynamics within a finite time interval, $\textnormal{NM}_1$.} The set of all quantum maps is denoted by a grey colored spheroid. All the free operations corresponding to quantum resource, say $\mathcal{K}$, are represented by a yellow colored region. The diameter of the set of these free operations, $\mathcal{D}(\Lambda)$, is denoted by $d$, and let's say that the quantum map $\Lambda$ for a finite time interval $\lbrack t, t+\tau \rbrack$ is represented by a black circle with $t, \tau \geq 0$. The minimum distance from the set of free operations for quantum map $\Lambda$ in between the time interval $\lbrack t, t+\tau \rbrack$ is denoted by an orange colored solid line. This solid line schematically represents $\textnormal{NM}_1 (\Lambda, t, \tau)$ for the quantum map $\Lambda$ within the time interval $\lbrack t, t+\tau \rbrack$.}
\label{fig1} 
\end{figure}

Furthermore, we establish another measure or quantifier of the resourcefulness of the quantum resource, $\mathcal{K}$, for any quantum map $\Lambda$ in between the time interval $[t, t + \tau]$ with $t$, $\tau \geq 0$ and it will be 
\begin{align} 
\label{eqn-NM2} 
&\textnormal{NM}_2 (\Lambda, t, \tau) \coloneqq \nonumber \\ &\min_{\Lambda_{\mathcal{K}} \in \mathcal{D}(\Lambda)} T \left[ (\Lambda (t+\tau,t) \otimes \mathbbm{1}) \ket{\phi^{+}} \bra{\phi^{+}} || (\Lambda_{\mathcal{K}} \otimes \mathbbm{1}) \ket{\phi^{+}} \bra{\phi^{+}} \right]. 
\end{align} 
In the above expression, $\ket{\phi^{+}} \bra{\phi^{+}}$ 
represents the maximally entangled state between the system and environment, and minimization is taken over all free operations $\Lambda_{\mathcal{K}}$ in the resource theory of quantum resource $\mathcal{K}$. $T[\rho_1||\rho_2] $ denotes any distance measure between the states $\rho_1$ and $\rho_2$.

\section{Measures of resourcefulness of any quantum resource and Measures of indivisibility for quantum channel: Understanding the Connection}
\label{sec3}
In this section, we will try to find connections between the measures of resourcefulness of any quantum resource for any quantum channel and the measures of indivisibility of the channel, where the indivisibility of the channel can be P-indivisibility or CP-indivisibility. We will use either $\textnormal{NM}_1 (\Lambda, t, \tau)$ or $\textnormal{NM}_2 (\Lambda, t, \tau)$ to measure the resourcefulness of the quantum resource, $\mathcal{K}$, for the quantum map $\Lambda$ in between the time interval $\lbrack t, t+\tau \rbrack$ with $t, \tau \geq 0$.

\subsection{Relationship between the measure of resourcefulness of a quantum resource for any quantum channel and its measure of P-indivisibility}

Here, our goal is to connect the measure of resourcefulness of any quantum resource for the quantum map $\Lambda$, $\textnormal{NM}_1 (\Lambda, t, \tau)$, and the measure of P-indivisibility, $\textnormal{P-I} (\Lambda, t, \tau)$, of the channel within the time interval $\lbrack t, t+\tau \rbrack$ where $t, \tau \geq 0$. In this subsection, we consider trace distance as a distance measure between any two quantum states, and by ``$\rho$", we mean that the state of the system is $\rho$ at initial time. Let us assume that $\Lambda_{\mathcal{K}}^j$ is the free operation in the resource theory of quantum resource $\mathcal{K}$ for which $D \lbrack \Lambda (t+\tau,t) \Lambda (t,0) \rho_j || \Lambda_{\mathcal{K}} (t+\tau,t) \Lambda (t,0) \rho_j \rbrack$ gets minimum over all other free operations when the initial state of the system is $\rho_j$ where $j$ denotes different initial states of the system. It is evident that for every initial state of the system $\rho_j$ on the Hilbert space $\mathcal{H}_S$, there will be at least one such free operation $\Lambda_{\mathcal{K}}^j$.
We will have
\begin{align}
\label{P-I-NM-eqn7}
    D \lbrack \Lambda (t+\tau,t) \Lambda (t,0) \rho_j || \Lambda_{\mathcal{K}}^j (t+\tau,t) \Lambda (t,0) \rho_j \rbrack \nonumber \\ \leq  \textnormal{NM}_1 (\Lambda, t, \tau),
\end{align}
which holds for any quantum state, $\rho_j$, of the system.

Then, we can write $D \lbrack \Lambda (t+\tau,0) \rho_1 ||\Lambda (t+\tau,0) \rho_2 \rbrack$ as follows
\begin{align}
\label{P-I-NM-eqn2}
&D \lbrack \Lambda (t+\tau,0) \rho_1 ||\Lambda (t+\tau,0) \rho_2 \rbrack \leq \nonumber \\ 
& D \lbrack \Lambda (t+\tau,0) \rho_1 || \Lambda_{\mathcal{K}}^1 (t+\tau,t) \Lambda (t,0) \rho_1 \rbrack + \nonumber \\ 
& D \lbrack \Lambda_{\mathcal{K}}^1 (t+\tau,t) \Lambda (t,0) \rho_1  || \Lambda (t+\tau,0) \rho_2 \rbrack,
\end{align}
where $\rho_1$ and $\rho_2$ are any two arbitrary initial states of the system. We have used the fact that trace distance follows the triangle inequality.
When we consider $\rho_1 \in \mathcal{D}(\mathcal{H}_S)$ as the initial state of the system, $ \Lambda_{\mathcal{K}}^1$ is the free operation for which $D \lbrack \Lambda (t+\tau,t) \Lambda (t,0) \rho_1 || \Lambda_{\mathcal{K}} (t+\tau,t) \Lambda (t,0) \rho_1 \rbrack$ gets minimum over all other free operations in the resource theory of quantum resource $\mathcal{K}$.
Now, we will use the symmetry property of trace distance in the last term of the previous inequality~\eqref{P-I-NM-eqn2} and we get
\begin{align}
&D \lbrack \Lambda (t+\tau,0) \rho_1 ||\Lambda (t+\tau,0) \rho_2 \rbrack \leq \nonumber \\ &D \lbrack \Lambda (t+\tau,0) \rho_1 ||  \Lambda_{\mathcal{K}}^1 (t+\tau,t) \Lambda (t,0) \rho_1 \rbrack + \nonumber \\ &D \lbrack \Lambda (t+\tau,0) \rho_2||  \Lambda_{\mathcal{K}}^1 (t+\tau,t) \Lambda (t,0) \rho_1 \rbrack. \label{P-I-NM-eqn3}
\end{align} 
Then, we can express $D \lbrack \Lambda (t+\tau,0) \rho_2|| \Lambda_{\mathcal{K}}^1 (t+\tau,t) \Lambda (t,0) \rho_1 \rbrack$ as 
\begin{align}
&D \lbrack \Lambda (t+\tau,0) \rho_2|| \Lambda_{\mathcal{K}}^1 (t+\tau,t) \Lambda (t,0) \rho_1 \rbrack \leq \nonumber \\ &D \lbrack \Lambda (t+\tau, 0) \rho_1 || \Lambda_{\mathcal{K}}^2 (t+\tau,t) \Lambda (t,0) \rho_2 \rbrack + \nonumber \\ &D \lbrack \Lambda_{\mathcal{K}}^2 (t+\tau,t) \Lambda (t,0) \rho_2 || \Lambda_{\mathcal{K}}^1 (t+\tau,t) \Lambda (t,0) \rho_1 \rbrack, \label{P-I-NM-eqn4} \\
&\Rightarrow D \lbrack \Lambda (t+\tau,0) \rho_2|| \Lambda_{\mathcal{K}}^1 (t+\tau,t) \Lambda (t,0) \rho_1 \rbrack \leq \nonumber \\ &D \lbrack \Lambda (t+\tau, 0) \rho_2 || \Lambda_{\mathcal{K}}^2 (t+\tau,t) \Lambda (t,0) \rho_2 \rbrack + \nonumber \\ &D \lbrack \Lambda_{\mathcal{K}}^1 (t+\tau,t) \Lambda (t,0) \rho_1 || \Lambda_{\mathcal{K}}^2 (t+\tau,t) \Lambda (t,0) \rho_2 \rbrack ,\label{P-I-NM-eqn5}
\end{align}
where $\Lambda_{\mathcal{K}}^2$ represents the free operation for which the quantity $D \lbrack \Lambda (t+\tau,t) \Lambda (t,0) \rho_2 || \Lambda_{\mathcal{K}} (t+\tau,t) \Lambda (t,0) \rho_2 \rbrack$ achieves its minimum value over all other free operations within the resource theory of quantum resource $\mathcal{K}$, when considering $\rho_2 \in \mathcal{D}(\mathcal{H}_S)$ as the initial state of the system.
We have used triangle inequality of distance measure and symmetric property of trace distance in the inequalities~\eqref{P-I-NM-eqn4} and~\eqref{P-I-NM-eqn5} respectively. 
Substituting $D \lbrack \Lambda (t+\tau,0) \rho_2|| \Lambda_{\mathcal{K}}^1 (t+\tau,t) \Lambda (t,0) \rho_1 \rbrack$ from~\eqref{P-I-NM-eqn5} into~\eqref{P-I-NM-eqn3}, we obtain

\begin{align}
    &D \lbrack \Lambda (t+\tau,0) \rho_1 || \Lambda (t+\tau,0) \rho_2 \rbrack \leq \nonumber \\
    & D \lbrack \Lambda (t+\tau,0) \rho_1 || \Lambda_{\mathcal{K}}^1  (t+\tau,t) \Lambda (t,0) \rho_1 \rbrack + \nonumber \\ 
    &D \lbrack \Lambda (t+\tau, 0) \rho_2 || \Lambda_{\mathcal{K}}^2 (t+\tau,t) \Lambda (t,0) \rho_2 \rbrack + \nonumber \\ 
    &D \lbrack \Lambda_{\mathcal{K}}^1 \Lambda (t,0) \rho_1 || \Lambda_{\mathcal{K}}^2 (t+\tau,t) \Lambda (t,0) \rho_2 \rbrack.\label{P-I-NM-eqn6}
\end{align}

We can say from~\eqref{P-I-NM-eqn7} that the first and second terms in the right hand side of the above inequality~\eqref{P-I-NM-eqn6} each is upper bounded by $\textnormal{NM}_1 (\Lambda, t, \tau)$.  Then, we can write~\eqref{P-I-NM-eqn6} as follows:

\begin{align}
    D \lbrack \Lambda (t+\tau,0) \rho_1 || \Lambda (t+\tau,0) \rho_2 \rbrack \leq 2 \textnormal {NM}_1(\Lambda,t,\tau) \phantom{abc} \nonumber \\ 
    + D \lbrack \Lambda_{\mathcal{K}}^1 (t+\tau,t) \Lambda (t,0) \rho_1 || \Lambda_{\mathcal{K}}^2 (t+\tau,t) \Lambda (t,0) \rho_2 \rbrack. \label{P-I-NM-eqn8}
\end{align}

Since, for any distance measure, the distance between any two states must be greater or equal to zero, then we can say that for any map $\Lambda$, $D\lbrack \Lambda(t,0) \rho_1||\Lambda(t,0) \rho_2 \rbrack \geq 0$ where $\rho_1 \in \mathcal{D}(\mathcal{H}_S)$ and $\rho_2 \in \mathcal{D}(\mathcal{H}_S)$ are any two arbitrary initial states of the system.
Therefore, we will have
\begin{widetext}
\begin{align}
\max_{\rho_1,\rho_2 \in \mathcal{D}(\mathcal{H}_S)} &\lbrace D \lbrack \Lambda (t+\tau,0) \rho_1 || \Lambda (t+\tau,0) \rho_2 \rbrack - D \lbrack \Lambda(t,0) \rho_1 || \Lambda(t,0) \rho_2 \rbrack \rbrace \leq \max_{\rho_1,\rho_2} \lbrace 2 \textnormal {NM}_1(\Lambda,t,\tau)  + \nonumber \\ &D\lbrack\Lambda_{\mathcal{K}}^1 (t+\tau,t) \Lambda(t,0) \rho_1 ||  \Lambda_{\mathcal{K}}^2 (t+\tau,t) \Lambda(t,0) \rho_2 \rbrack - D \lbrack \Lambda(t,0) \rho_1||\Lambda(t,0) \rho_2 \rbrack \rbrace. \label{P-I-NM-eqn9} \\
\Rightarrow \max_{\rho_1,\rho_2 \in  \mathcal{D}(\mathcal{H}_S)}&\lbrace D \lbrack \Lambda (t+\tau,0) \rho_1 || \Lambda (t+\tau,0) \rho_2 \rbrack - D \lbrack \Lambda(t,0) \rho_1 || \Lambda(t,0) \rho_2 \rbrack \rbrace \leq \textnormal{2NM}_1(\Lambda,t,\tau) + \nonumber \\ &\max_{\rho_1,\rho_2 \in  \mathcal{D}(\mathcal{H}_S)} D\lbrack\Lambda_{\mathcal{K}}^1 (t+\tau,0) \rho_1(t)||  \Lambda_{\mathcal{K}}^2 (t+\tau,0)  \rho_2(t)\rbrack - \min_{\rho_1,\rho_2 \in  \mathcal{D}(\mathcal{H}_S)} D\lbrack \rho_1(t)|| \rho_2(t)\rbrack, \label{P-I-NM-eqn10}\\
\Rightarrow \max_{\rho_1,\rho_2 \in  \mathcal{D}(\mathcal{H}_S)} & \lbrace D \lbrack \Lambda (t+\tau,0) \rho_1 || \Lambda (t+\tau,0) \rho_2 \rbrack - D \lbrack \Lambda(t,0) \rho_1 || \Lambda(t,0) \rho_2 \rbrack \rbrace \leq 2 \textnormal{NM}_1(\Lambda,t,\tau) + d, \label{P-I-NM-eqn11} 
\end{align}   
\end{widetext}

where $\rho_1(t) \coloneqq \Lambda(t,0) \rho_1$ and $\rho_2(t) \coloneqq \Lambda(t,0) \rho_2$.
In the~\eqref{P-I-NM-eqn10}, we have used the fact that $\max_{x} \lbrace f(x) + g(x) \rbrace \leq \max_x f(x) + \max_x g(x)$ for any functions $f(x), g(x)$. 
In the~\eqref{P-I-NM-eqn10}, we have used the fact that $\min_{\rho_1,\rho_2} D\lbrack \rho_1(t)|| \rho_2(t)\rbrack = 0$, 
and $d \coloneqq \max_{\rho_1,\rho_2} D\lbrack\Lambda_{\mathcal{K}}^1 (t+\tau,0) \rho_1(t)||  \Lambda_{\mathcal{K}}^2 (t+\tau,0) \rho_2(t)\rbrack$. 
Here, $d$ represents the diameter of the set consisting of all free operations in the framework of the resource theory of quantum resource $\mathcal{K}$. Then, we can have
\begin{align}
\textnormal{P-I}(\Lambda, t, \tau) &\leq  2 \textnormal{NM}_1(\Lambda,t,\tau) + d \label{P-I-NM-eqn12},
\end{align}
where we have got the left hand side of~\eqref{P-I-NM-eqn12} using the definition of quantifier of P-indivisibility of any quantum channel described in~\eqref{eqn-def-P-I}.

Thus, the inequality~\eqref{P-I-NM-eqn12} gives us a relation between a measure of resourcefulness of a quantum resource, $\textnormal{NM}_1 (\Lambda, t, \tau)$, and a measure of P-indivisibility, $\textnormal{P-I} (\Lambda, t, \tau)$, for any quantum channel $\Lambda (t , t + \tau)$ where $t, \tau \geq 0$. 
Since for a quantum resource $\mathcal{K}$ in between a finite time interval, $d$ is just a fixed value, i.e., it is independent of the quantum channel $\Lambda$, the inequality~\eqref{P-I-NM-eqn12} tells us that the measure of P-indivisibility, $\textnormal{P-I} (\Lambda, t, \tau)$, for any channel, $\Lambda (t , t + \tau)$, will give a lower bound of $\textnormal{NM}_1(\Lambda,t,\tau)$. 
In the framework of resource theory of quantum resource, $\mathcal{K}$, $\textnormal{NM}_1(\Lambda_{\mathcal{K}},t,\tau) = 0 $ for any free operation $\Lambda_{\mathcal{K}} \in \mathcal{D}(\Lambda)$. Hence, from~\eqref{P-I-NM-eqn12}, we can also say that $\textnormal{P-I}(\Lambda_{\mathcal{K}}, t, \tau) \leq d$ for any free operation, $\Lambda_{\mathcal{K}}$, in between the time interval $\lbrack t, t+\tau \rbrack$ with $t, \tau \geq 0$.

\subsection{Relationship between the measure of resourcefulness of a quantum resource for any quantum channel and its measure of CP-indivisibility}

Here, we will try to find a relation between the measure of resourcefulness of any quantum resource, $\textnormal{NM}_2 (\Lambda, t, \tau)$, for any quantum channel $\Lambda (t , t + \tau)$ and the quantifier of CP-indivisibility of the channel, $\textnormal{CP-I}(\Lambda,t,\tau)$ with $t, \tau \geq 0$.
Let us assume that $\Lambda_{\mathcal{K'}}$ is the free operation of the resource theory of the quantum resource $\mathcal{K}$ for which
$D \lbrack (\Lambda (t+\tau,t) \otimes \mathbbm{1}) \ket{\phi^{+}} \bra{\phi^{+}} || (\Lambda_{\mathcal{K'}} \otimes \mathbbm{1}) \ket{\phi^{+}} \bra{\phi^{+}} \rbrack$ gets minimum over all other free operations. Then Eq.~\eqref{eqn-NM2} can be written as
\begin{align}
 \textnormal{NM}_2 (\Lambda, t, \tau) \phantom{ghum peyechhe, baRi ja, ghor ja, alu peyaj poTol ab}\nonumber \\ 
  = D \lbrack (\Lambda (t+\tau,t) \otimes \mathbbm{1}) \ket{\phi^{+}} \bra{\phi^{+}} || (\Lambda_{\mathcal{K'}} \otimes \mathbbm{1}) \ket{\phi^{+}} \bra{\phi^{+}} \rbrack, \phantom{abarcd}\\
= \frac{1}{2} || (\Lambda (t+\tau,t) \otimes \mathbbm{1}) \ket{\phi^{+}} \bra{\phi^{+}} - (\Lambda_{\mathcal{K'}} \otimes \mathbbm{1}) \ket{\phi^{+}} \bra{\phi^{+}}||_1, \phantom{abc}\\
 \leq \frac{1}{2} || (\Lambda (t+\tau,t) \otimes \mathbbm{1}) \ket{\phi^{+}} \bra{\phi^{+}} ||_1  \phantom{Subhaschandra Chittar}\nonumber \\ 
 +\frac{1}{2} ||(\Lambda_{\mathcal{K'}} \otimes \mathbbm{1}) \ket{\phi^{+}} \bra{\phi^{+}}||_1, \quad \quad \quad \label{CP-I-NM-eqn1}
\end{align}
where $||A||_1$ denotes the trace norm for matrix A and we have used trace distance as a distance measure in the definition of $\textnormal{NM}_2 (\Lambda, t, \tau)$. 
In~\eqref{CP-I-NM-eqn1}, we have used the fact that the trace norm follows the triangular property, $ ||A+B||_1 \leq ||A||_1 + ||B||_1 $ for any two matrices $A$ and $B$.
All the free operations in the resource theory of quantum resource, $\mathcal{K}$, can be divided into CP-divisible and CP-indivisible maps. From the definitions of trace norm, we can say that $||(\Lambda_{\mathcal{K}} \otimes \mathbbm{1}) \ket{\phi^{+}} \bra{\phi^{+}}||_1 > 1$ for all CP-indivisible free operations, since the operations are trace preserving. 
Now we assume that among all the free operations, there exists at least one CP-divisible quantum free operation for all finite time interval.
Hence, $||(\Lambda_{\mathcal{K'}} \otimes \mathbbm{1}) \ket{\phi^{+}} \bra{\phi^{+}}||_1 = 1$.
Then, from~\eqref{CP-I-NM-eqn1}, we can write
\begin{align}
\textnormal{NM}_2 (\Lambda, t, \tau) &\leq  \frac{1}{2} \textnormal{CP-I} (\Lambda, t, \tau) + 1 \label{CP-I-NM-eqn3}.
\end{align}
In~\eqref{CP-I-NM-eqn3}, we have used the mathematical form of the quantifier of CP-indivisibility for any map $\Lambda$ within the time interval $\lbrack t , t+ \tau \rbrack$ described in~\eqref{eqn-def-CP-I} where $t, \tau \geq 0$.

Thus, the measure of resourcefulness of any quantum resource, $\textnormal{NM}_2 (\Lambda, t, \tau)$, is related with the measure of CP-indivisibility, $\textnormal{CP-I} (\Lambda, t, \tau)$, for any quantum channel $\Lambda$ in between any finite time interval $\lbrack t, t+ \tau \rbrack$ where $t, \tau \geq 0$. From the above inequality, we can also say that we will be able to tell about an upper bound of a measure of resourcefulness of any quantum resource, $\textnormal{NM}_2 (\Lambda, t, \tau)$, if we know the value of measure of CP-indivisibility, $\textnormal{CP-I} (\Lambda, t, \tau)$, for any quantum channel within any finite time interval. Since $\textnormal{CP-I} (\Lambda, t, \tau) = 0$ for any CP-divisible quantum map, we can say from~\eqref{CP-I-NM-eqn3} that the measure of resourcefulness of a quantum resource for any CP-divisible quantum map will show, $\textnormal{NM}_2 (\Lambda, t, \tau) \leq 1$, for any finite time interval.

\begin{figure}[h!]
\includegraphics[scale=1.03]{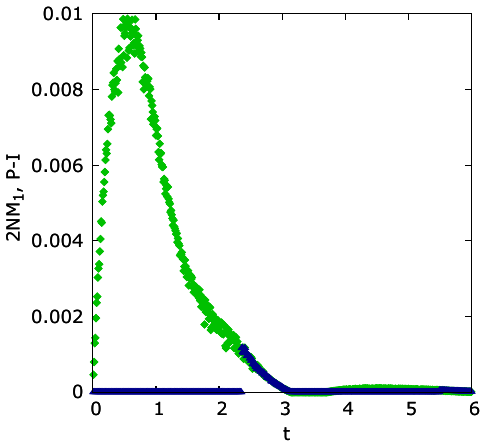}
\caption{\textbf{Verification of the relationship between the measure of resourcefulness of quantum resource $\textnormal{NM}_1 (\Lambda, t, \tau)$, and the measure of P-indivisibility, $\textnormal{P-I} (\Lambda, t, \tau)$, for a particular quantum dynamics within any finite time interval.} Here, we consider ``quantum non-Markovianity" as the quantum resource. The quantum dynamics considered
is the dynamics generated by the Jaynes-Cummings model at resonance parameterized by $\gamma_0 = 2$ and $\lambda =2$ within any time $t$ to $t+\tau$ where $t \geq 0$ and $\tau = 0.01$.
We plot $2 \textnormal{NM}_1 (\Lambda, t, \tau)$ and $\textnormal{P-I} (\Lambda, t, \tau)$ along the vertical axis while time along the vertical axis for the quantum dynamics.
The green colored dotted curve and blue colored dotted curve represent $2 \textnormal{NM}_1 (\Lambda, t, \tau)$ and $\textnormal{P-I} (\Lambda, t, \tau)$, respectively, for the particular quantum dynamics represented by Eqs.~\eqref{eqn-Linblad-JC} and~\eqref{eqn-JC-decay} with $\lambda = \gamma_0 = 2$ within any time interval. We can see that $2 \textnormal{NM}_1 (\Lambda, t, \tau)$ is always greater or equal to $\textnormal{P-I} (\Lambda, t, \tau)$. Thus, the inequality~\eqref{P-I-NM-eqn12} is satisfied for the particular dynamics within any time interval. All the quantities are dimensionless.}
\label{fig2} 
\end{figure}

\section{Validation of the relations using quantum non-Markovianity as channel resource}
\label{sec4}
In this section, we will present an example to validate our two primary findings. One is a relationship between the measure of resourcefulness of any quantum resource, $\textnormal{NM}_1  (\Lambda, t, \tau)$, and the measure of P-indivisibility, $\textnormal{P-I} (\Lambda, t, \tau)$, for any quantum channel $\Lambda (t, t+\tau)$, represented in~\eqref{P-I-NM-eqn12}. And another is a relationship between the measure of resourcefulness of any quantum resource, $\textnormal{NM}_2  (\Lambda, t, \tau)$, and the measure of CP-indivisibility, $\textnormal{CP-I}  (\Lambda, t, \tau)$, for any quantum channel $\Lambda (t, t+\tau)$, shown in~\eqref{CP-I-NM-eqn3}.  But, we will check our results for quantum resource ``quantum non-Markovianity".

We will assume here that the system is in the Hilbert space $\mathbb{C}^2$ and 
the local Hamiltonian of the single-qubit system is, $H_S = \omega_0 \sigma_z$ where $\omega_0$ is the transition frequency of the system. The two-level system is interacting with a bath of harmonic oscillators in the vacuum state. The local Hamiltonian of the environment is, $H_E = \sum_k \omega_k b_k^\dagger b_k$
where $b_k^\dagger$ and $b_k$ denotes the raising and lowering operators of $k$-th mode with frequency $\omega_k$. The interacting Hamiltonian is, $H_{SE} = \sigma_{+} \otimes B + \sigma_{-} \otimes B^\dagger$ where $B \coloneqq \sum_k g_k b_k$ and $g_k$ denotes the coupling constants. This model will be called Jaynes-Cummings model at resonance~\cite{petru-book,lidar-notes} if 
the corresponding Lindblad master equation of system dynamics is given by

\begin{align}
\label{eqn-Linblad-JC}
\dot{\rho} = \gamma (t) \mathcal{L} (\rho(t)),    
\end{align}
for all times other than a few that depend on the parameters of decay rate, $\gamma_0$ and $\lambda$.
Here, the Lindblad is $\mathcal{L}(\rho(t)) = \sigma_{-} \rho \sigma_{+} - \frac{1}{2} \lbrace \sigma_{+} \sigma_{-} , \rho(t) \rbrace$, and the time-dependent decay rate is
\begin{align}
\label{eqn-JC-decay}
\gamma (t) = \frac{2 \lambda \gamma_0 \sinh{(\frac{\delta t}{2}})}{\delta \cosh{(\frac{\delta t}{2})}+\lambda \sinh{(\frac{\delta t}{2}})}; \hspace{4 mm} \delta \coloneqq \sqrt{\lambda^2 - 2 \gamma_0 \lambda},
\end{align}
where $\gamma_0$ and $\lambda$ are positive constants. $\sigma_{+}$ and $\sigma_{-}$ are the two eigenstates of $\sigma_x$ with eigenvalues $1$ and $-1$ respectively.
The Lindblad equation can describe the dynamics of the system at all times when $\gamma_0 < \frac{\lambda}{2}$. However, when $\gamma_0 > \frac{\lambda}{2}$, the Lindblad equation is valid for all times except for a few, at which the system reaches the ground eigenstate of the local Hamiltonian of the system.
When $\gamma_0 < \frac{\lambda}{2}$, the quantum dynamics for the Jaynes-Cummings model at resonance exhibits Markovian dynamics, and when $\gamma_0 > \frac{\lambda}{2}$, it demonstrates non-Markovian dynamics.
In this context, Markovian dynamics refer to a scenario where the population at the excited eigenstate of the local Hamiltonian of the two-level system decreases monotonically over time. Conversely, if the population at the excited state exhibits any non-monotonic behavior over time, we characterize it as non-Markovian dynamics.
\begin{figure}[h!]
\includegraphics[scale=1.08]{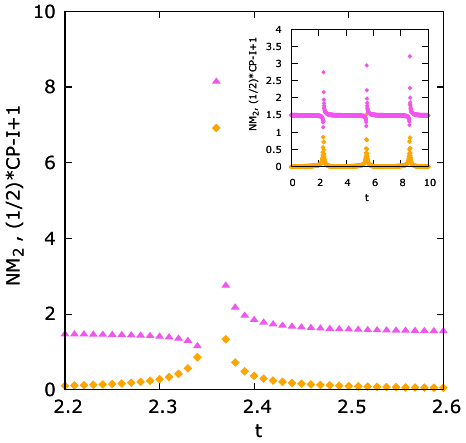}
\caption{\textbf{Verification of the relationship between the measure of resourcefulness of ``quantum non-Markovianity", $\textnormal{NM}_2 (\Lambda, t, \tau)$, and a measure of CP-indivisibility, $\textnormal{CP-I} (\Lambda, t, \tau)$, for a particular quantum dynamics within any finite time interval.}
We plot time (t) along the horizontal axis and $\textnormal{NM}_2(\Lambda, t, \tau)$ and $\frac{1}{2}\textnormal{CP-I}(\Lambda, t, \tau)+1$ along the vertical axis for the particular quantum dynamics represented by Eqs.~\eqref{eqn-Linblad-JC} and~\eqref{eqn-JC-decay} with $\lambda = \gamma_0 = 2$ within any finite time interval $\lbrack t, t+\tau \rbrack$ choosing $\tau = 0.01$ and $t \geq 0$.
The orange colored dotted curve and magenta colored dotted curve represent $\textnormal{NM}_2 (\Lambda, t, \tau)$ and $\frac{1}{2}\textnormal{CP-I}(\Lambda, t, \tau) + 1$, respectively. We can observe that the inequality~\eqref{P-I-NM-eqn12} is satisfied for the particular dynamics within any finite time interval. All the quantities are dimensionless.}
\label{fig3} 
\end{figure}

Here, we will verify the relationships between measures of resourcefulness of quantum resource ``quantum non-Markovianity" and the measures of indivisibility for the quantum dynamics generated by the Jaynes-Cummings model at resonance parameterized by $\gamma_0 = 2$ and $\lambda =2$ within any finite time interval.
While calculating the measures of resourcefulness of quantum resource ``quantum non-Markovianity", $\textnormal{NM}_1 (\Lambda, t, \tau)$ and $\textnormal{NM}_2 (\Lambda, t, \tau)$ for the quantum dynamics generated by the Jaynes-Cummings model at resonance parameterized by $\gamma_0 = 2$ and $\lambda =2$ within any finite time interval, note that all the evolutions generated by the Jaynes-Cummings model at resonance parameterized by $\gamma_0$ and $\lambda$ following $\gamma_0 < \frac{\lambda}{2}$ will act as free operations here. 
To compute $\textnormal{NM}_1 (\Lambda, t, \tau)$ and $\textnormal{NM}_2 (\Lambda, t, \tau)$, however, we shall only optimize over those Markovian maps that have $\lambda = 2$ and $0 < \gamma_0 < 1$. We choose $\tau = 0.01$ to evaluate $\textnormal{NM}_1 (\Lambda, t, \tau)$, $\textnormal{NM}_2 (\Lambda, t, \tau)$, $\textnormal{P-I} (\Lambda, t, \tau)$, and $\textnormal{CP-I} (\Lambda, t, \tau)$ for the quantum dynamics generated by the Jaynes-Cummings model at resonance parameterized by $\gamma_0 = 2$ and $\lambda =2$ within any time $t$ to $t+\tau$ where $t \geq 0$.

In Fig. \ref{fig2}, we show the behavior of $2 \textnormal{NM}_1 (\Lambda, t, \tau)$ and $\textnormal{P-I} (\Lambda, t, \tau)$ by the green and blue colored dotted curves, respectively, for the particular dynamics within any finite time interval. We can see that the inequality~\eqref{P-I-NM-eqn12} is satisfied by the quantum dynamics for all finite time intervals. Next, we plot $\textnormal{NM}_2 (\Lambda, t, \tau)$ and $\frac{1}{2}\textnormal{CP-I} (\Lambda, t, \tau) +1$ for the quantum dynamics described above within any finite time interval, represented by an orange and magenta colored dotted curves in Fig. \ref{fig3}. We can see that the inequality~\eqref{CP-I-NM-eqn3} is satisfied in this scenario for all time intervals and verifies our second inequality connecting the measure of resourcefulness of quantum resource ``quantum non-Markovianity", $\textnormal{NM}_2 (\Lambda, t, \tau)$, and measure of CP-indivisibility, $\textnormal{CP-I} (\Lambda, t, \tau)$, for the quantum dynamics within any finite time interval.

\section{Conclusions}
\label{sec5}
Studying quantum systems in presence of their environments is important from a fundamental perspective, for example for a better understanding of quantum thermodynamics,  as well as from the viewpoint of applications, for example to obtain a better management of noise in quantum technologies. 
Quantum channels are often categorized according to their divisibilities, and we 
 proposed P- and CP-indivisibility measures for any quantum channel over any finite time interval.
We went on to quantify - in two ways - the resourcefulness of a quantum channel with respect to an arbitrary channel resource, again assuming that the channel acts for a finite time interval.

The aim of this work was a find a bridge between the measures of indivisibilities of quantum channels with their resourcefulness with respect to an arbitrary given channel resource. We found two distinct relations between the two ends of that envisaged bridge, in the form of inequalities between the measures of those two areas. 
Finally, we validated our findings through a numerical-handled example involving the Jaynes-Cummings model at resonance, considering quantum non-Markovianity as the quantum channel  resource.

\bibliography{nm_qm}

%merlin.mbs apsrev4-1.bst 2010-07-25 4.21a (PWD, AO, DPC) hacked
%Control: key (0)
%Control: author (0) dotless jnrlst
%Control: editor formatted (1) identically to author
%Control: production of article title (0) allowed
%Control: page (1) range
%Control: year (0) verbatim
%Control: production of eprint (0) enabled
\begin{thebibliography}{47}%
\makeatletter
\providecommand \@ifxundefined [1]{%
 \@ifx{#1\undefined}
}%
\providecommand \@ifnum [1]{%
 \ifnum #1\expandafter \@firstoftwo
 \else \expandafter \@secondoftwo
 \fi
}%
\providecommand \@ifx [1]{%
 \ifx #1\expandafter \@firstoftwo
 \else \expandafter \@secondoftwo
 \fi
}%
\providecommand \natexlab [1]{#1}%
\providecommand \enquote  [1]{``#1''}%
\providecommand \bibnamefont  [1]{#1}%
\providecommand \bibfnamefont [1]{#1}%
\providecommand \citenamefont [1]{#1}%
\providecommand \href@noop [0]{\@secondoftwo}%
\providecommand \href [0]{\begingroup \@sanitize@url \@href}%
\providecommand \@href[1]{\@@startlink{#1}\@@href}%
\providecommand \@@href[1]{\endgroup#1\@@endlink}%
\providecommand \@sanitize@url [0]{\catcode `\\12\catcode `\$12\catcode `\&12\catcode `\#12\catcode `\^12\catcode `\_12\catcode `\%12\relax}%
\providecommand \@@startlink[1]{}%
\providecommand \@@endlink[0]{}%
\providecommand \url  [0]{\begingroup\@sanitize@url \@url }%
\providecommand \@url [1]{\endgroup\@href {#1}{\urlprefix }}%
\providecommand \urlprefix  [0]{URL }%
\providecommand \Eprint [0]{\href }%
\providecommand \doibase [0]{http://dx.doi.org/}%
\providecommand \selectlanguage [0]{\@gobble}%
\providecommand \bibinfo  [0]{\@secondoftwo}%
\providecommand \bibfield  [0]{\@secondoftwo}%
\providecommand \translation [1]{[#1]}%
\providecommand \BibitemOpen [0]{}%
\providecommand \bibitemStop [0]{}%
\providecommand \bibitemNoStop [0]{.\EOS\space}%
\providecommand \EOS [0]{\spacefactor3000\relax}%
\providecommand \BibitemShut  [1]{\csname bibitem#1\endcsname}%
\let\auto@bib@innerbib\@empty
%</preamble>
\bibitem [{\citenamefont {Rivas}\ and\ \citenamefont {Huelga}(2012)}]{Rivas-open}%
  \BibitemOpen
  \bibfield  {author} {\bibinfo {author} {\bibfnamefont {{\'{A}}}~\bibnamefont {Rivas}}\ and\ \bibinfo {author} {\bibfnamefont {S.~F.}\ \bibnamefont {Huelga}},\ }\href {https://doi.org/10.1007%2F978-3-642-23354-8} {\emph {\bibinfo {title} {Open Quantum Systems}}}\ (\bibinfo  {publisher} {Springer Berlin Heidelberg},\ \bibinfo {year} {2012})\BibitemShut {NoStop}%
\bibitem [{\citenamefont {Rivas}\ \emph {et~al.}(2014)\citenamefont {Rivas}, \citenamefont {Huelga},\ and\ \citenamefont {Plenio}}]{RHP-review}%
  \BibitemOpen
  \bibfield  {author} {\bibinfo {author} {\bibfnamefont {{\'{A}}}~\bibnamefont {Rivas}}, \bibinfo {author} {\bibfnamefont {S.~F.}\ \bibnamefont {Huelga}}, \ and\ \bibinfo {author} {\bibfnamefont {M.~B.}\ \bibnamefont {Plenio}},\ }\bibfield  {title} {\enquote {\bibinfo {title} {Quantum non-{M}arkovianity: characterization, quantification and detection},}\ }\href {https://doi.org/10.1088/0034-4885/77/9/094001} {\bibfield  {journal} {\bibinfo  {journal} {Rep. Prog. Phys.}\ }\textbf {\bibinfo {volume} {77}},\ \bibinfo {pages} {094001} (\bibinfo {year} {2014})}\BibitemShut {NoStop}%
\bibitem [{\citenamefont {Breuer}\ \emph {et~al.}(2016)\citenamefont {Breuer}, \citenamefont {Laine}, \citenamefont {Piilo},\ and\ \citenamefont {Vacchini}}]{BLPV-review}%
  \BibitemOpen
  \bibfield  {author} {\bibinfo {author} {\bibfnamefont {H-P.}\ \bibnamefont {Breuer}}, \bibinfo {author} {\bibfnamefont {E-M.}\ \bibnamefont {Laine}}, \bibinfo {author} {\bibfnamefont {J.}~\bibnamefont {Piilo}}, \ and\ \bibinfo {author} {\bibfnamefont {B.}~\bibnamefont {Vacchini}},\ }\bibfield  {title} {\enquote {\bibinfo {title} {Colloquium: Non-{M}arkovian dynamics in open quantum systems},}\ }\href {https://link.aps.org/doi/10.1103/RevModPhys.88.021002} {\bibfield  {journal} {\bibinfo  {journal} {Rev. Mod. Phys.}\ }\textbf {\bibinfo {volume} {88}},\ \bibinfo {pages} {021002} (\bibinfo {year} {2016})}\BibitemShut {NoStop}%
\bibitem [{\citenamefont {de~Vega}\ and\ \citenamefont {Alonso}(2017)}]{VA-review}%
  \BibitemOpen
  \bibfield  {author} {\bibinfo {author} {\bibfnamefont {I.}~\bibnamefont {de~Vega}}\ and\ \bibinfo {author} {\bibfnamefont {D.}~\bibnamefont {Alonso}},\ }\bibfield  {title} {\enquote {\bibinfo {title} {Dynamics of non-{M}arkovian open quantum systems},}\ }\href {https://link.aps.org/doi/10.1103/RevModPhys.89.015001} {\bibfield  {journal} {\bibinfo  {journal} {Rev. Mod. Phys.}\ }\textbf {\bibinfo {volume} {89}},\ \bibinfo {pages} {015001} (\bibinfo {year} {2017})}\BibitemShut {NoStop}%
\bibitem [{\citenamefont {Rivas}\ \emph {et~al.}(2010)\citenamefont {Rivas}, \citenamefont {Huelga},\ and\ \citenamefont {Plenio}}]{RHP-1}%
  \BibitemOpen
  \bibfield  {author} {\bibinfo {author} {\bibfnamefont {\'A.}\ \bibnamefont {Rivas}}, \bibinfo {author} {\bibfnamefont {S.~F.}\ \bibnamefont {Huelga}}, \ and\ \bibinfo {author} {\bibfnamefont {M.~B.}\ \bibnamefont {Plenio}},\ }\bibfield  {title} {\enquote {\bibinfo {title} {Entanglement and non-{M}arkovianity of quantum evolutions},}\ }\href {https://link.aps.org/doi/10.1103/PhysRevLett.105.050403} {\bibfield  {journal} {\bibinfo  {journal} {Phys. Rev. Lett.}\ }\textbf {\bibinfo {volume} {105}},\ \bibinfo {pages} {050403} (\bibinfo {year} {2010})}\BibitemShut {NoStop}%
\bibitem [{\citenamefont {Breuer}\ \emph {et~al.}(2009)\citenamefont {Breuer}, \citenamefont {Laine},\ and\ \citenamefont {Piilo}}]{BLP-1}%
  \BibitemOpen
  \bibfield  {author} {\bibinfo {author} {\bibfnamefont {H-P.}\ \bibnamefont {Breuer}}, \bibinfo {author} {\bibfnamefont {E-M.}\ \bibnamefont {Laine}}, \ and\ \bibinfo {author} {\bibfnamefont {J.}~\bibnamefont {Piilo}},\ }\bibfield  {title} {\enquote {\bibinfo {title} {Measure for the degree of non-{M}arkovian behavior of quantum processes in open systems},}\ }\href {https://link.aps.org/doi/10.1103/PhysRevLett.103.210401} {\bibfield  {journal} {\bibinfo  {journal} {Phys. Rev. Lett.}\ }\textbf {\bibinfo {volume} {103}},\ \bibinfo {pages} {210401} (\bibinfo {year} {2009})}\BibitemShut {NoStop}%
\bibitem [{\citenamefont {Breuer}\ and\ \citenamefont {Petruccione}(2002)}]{petru-book}%
  \BibitemOpen
  \bibfield  {author} {\bibinfo {author} {\bibfnamefont {H.~P.}\ \bibnamefont {Breuer}}\ and\ \bibinfo {author} {\bibfnamefont {F.}~\bibnamefont {Petruccione}},\ }\href@noop {} {\emph {\bibinfo {title} {The theory of open quantum systems}}}\ (\bibinfo  {publisher} {Oxford University Press},\ \bibinfo {address} {Great Clarendon Street},\ \bibinfo {year} {2002})\BibitemShut {NoStop}%
\bibitem [{\citenamefont {Wolf}\ \emph {et~al.}(2008)\citenamefont {Wolf}, \citenamefont {Eisert}, \citenamefont {Cubitt},\ and\ \citenamefont {Cirac}}]{marko-measure-1}%
  \BibitemOpen
  \bibfield  {author} {\bibinfo {author} {\bibfnamefont {M.~M.}\ \bibnamefont {Wolf}}, \bibinfo {author} {\bibfnamefont {J.}~\bibnamefont {Eisert}}, \bibinfo {author} {\bibfnamefont {T.~S.}\ \bibnamefont {Cubitt}}, \ and\ \bibinfo {author} {\bibfnamefont {J.~I.}\ \bibnamefont {Cirac}},\ }\bibfield  {title} {\enquote {\bibinfo {title} {Assessing non-{M}arkovian quantum dynamics},}\ }\href {https://link.aps.org/doi/10.1103/PhysRevLett.101.150402} {\bibfield  {journal} {\bibinfo  {journal} {Phys. Rev. Lett.}\ }\textbf {\bibinfo {volume} {101}},\ \bibinfo {pages} {150402} (\bibinfo {year} {2008})}\BibitemShut {NoStop}%
\bibitem [{\citenamefont {Lu}\ \emph {et~al.}(2010)\citenamefont {Lu}, \citenamefont {Wang},\ and\ \citenamefont {Sun}}]{marko-measure-4}%
  \BibitemOpen
  \bibfield  {author} {\bibinfo {author} {\bibfnamefont {X.}~\bibnamefont {Lu}}, \bibinfo {author} {\bibfnamefont {X.}~\bibnamefont {Wang}}, \ and\ \bibinfo {author} {\bibfnamefont {C.~P.}\ \bibnamefont {Sun}},\ }\bibfield  {title} {\enquote {\bibinfo {title} {Quantum fisher information flow and non-{M}arkovian processes of open systems},}\ }\href {https://link.aps.org/doi/10.1103/PhysRevA.82.042103} {\bibfield  {journal} {\bibinfo  {journal} {Phys. Rev. A}\ }\textbf {\bibinfo {volume} {82}},\ \bibinfo {pages} {042103} (\bibinfo {year} {2010})}\BibitemShut {NoStop}%
\bibitem [{\citenamefont {Hou}\ \emph {et~al.}(2011)\citenamefont {Hou}, \citenamefont {Yi}, \citenamefont {Yu},\ and\ \citenamefont {Oh}}]{marko-measure-5}%
  \BibitemOpen
  \bibfield  {author} {\bibinfo {author} {\bibfnamefont {S.~C.}\ \bibnamefont {Hou}}, \bibinfo {author} {\bibfnamefont {X.~X.}\ \bibnamefont {Yi}}, \bibinfo {author} {\bibfnamefont {S.~X.}\ \bibnamefont {Yu}}, \ and\ \bibinfo {author} {\bibfnamefont {C.~H.}\ \bibnamefont {Oh}},\ }\bibfield  {title} {\enquote {\bibinfo {title} {Alternative non-{M}arkovianity measure by divisibility of dynamical maps},}\ }\href {https://link.aps.org/doi/10.1103/PhysRevA.83.062115} {\bibfield  {journal} {\bibinfo  {journal} {Phys. Rev. A}\ }\textbf {\bibinfo {volume} {83}},\ \bibinfo {pages} {062115} (\bibinfo {year} {2011})}\BibitemShut {NoStop}%
\bibitem [{\citenamefont {Usha~Devi}\ \emph {et~al.}(2011)\citenamefont {Usha~Devi}, \citenamefont {Rajagopal},\ and\ \citenamefont {Sudha}}]{marko-measure-6}%
  \BibitemOpen
  \bibfield  {author} {\bibinfo {author} {\bibfnamefont {A.~R.}\ \bibnamefont {Usha~Devi}}, \bibinfo {author} {\bibfnamefont {A.~K.}\ \bibnamefont {Rajagopal}}, \ and\ \bibinfo {author} {\bibnamefont {Sudha}},\ }\bibfield  {title} {\enquote {\bibinfo {title} {Open-system quantum dynamics with correlated initial states, not completely positive maps, and non-{M}arkovianity},}\ }\href {https://link.aps.org/doi/10.1103/PhysRevA.83.022109} {\bibfield  {journal} {\bibinfo  {journal} {Phys. Rev. A}\ }\textbf {\bibinfo {volume} {83}},\ \bibinfo {pages} {022109} (\bibinfo {year} {2011})}\BibitemShut {NoStop}%
\bibitem [{\citenamefont {Devi}\ \emph {et~al.}(2012)\citenamefont {Devi}, \citenamefont {Rajagopal}, \citenamefont {Shenoy},\ and\ \citenamefont {Rendell}}]{marko-measure-7}%
  \BibitemOpen
  \bibfield  {author} {\bibinfo {author} {\bibfnamefont {A.}~\bibnamefont {Devi}}, \bibinfo {author} {\bibfnamefont {A.}~\bibnamefont {Rajagopal}}, \bibinfo {author} {\bibfnamefont {S.}~\bibnamefont {Shenoy}}, \ and\ \bibinfo {author} {\bibfnamefont {R.}~\bibnamefont {Rendell}},\ }\bibfield  {title} {\enquote {\bibinfo {title} {Interplay of quantum stochastic and dynamical maps to discern {M}arkovian and non-{M}arkovian transitions},}\ }\href {https://doi.org/10.4236/jqis.2012.23009} {\bibfield  {journal} {\bibinfo  {journal} {J. Quantum Inf. Sci}\ }\textbf {\bibinfo {volume} {2}},\ \bibinfo {pages} {47} (\bibinfo {year} {2012})}\BibitemShut {NoStop}%
\bibitem [{\citenamefont {Mazzola}\ \emph {et~al.}(2012)\citenamefont {Mazzola}, \citenamefont {Rodr\'{\i}guez-Rosario}, \citenamefont {Modi},\ and\ \citenamefont {Paternostro}}]{marko-measure-8}%
  \BibitemOpen
  \bibfield  {author} {\bibinfo {author} {\bibfnamefont {L.}~\bibnamefont {Mazzola}}, \bibinfo {author} {\bibfnamefont {C.~A.}\ \bibnamefont {Rodr\'{\i}guez-Rosario}}, \bibinfo {author} {\bibfnamefont {K.}~\bibnamefont {Modi}}, \ and\ \bibinfo {author} {\bibfnamefont {M.}~\bibnamefont {Paternostro}},\ }\bibfield  {title} {\enquote {\bibinfo {title} {Dynamical role of system-environment correlations in non-{M}arkovian dynamics},}\ }\href {https://link.aps.org/doi/10.1103/PhysRevA.86.010102} {\bibfield  {journal} {\bibinfo  {journal} {Phys. Rev. A}\ }\textbf {\bibinfo {volume} {86}},\ \bibinfo {pages} {010102} (\bibinfo {year} {2012})}\BibitemShut {NoStop}%
\bibitem [{\citenamefont {Rodríguez-Rosario}\ \emph {et~al.}(2012)\citenamefont {Rodríguez-Rosario}, \citenamefont {Modi}, \citenamefont {Mazzola},\ and\ \citenamefont {Aspuru-Guzik}}]{marko-measure-9}%
  \BibitemOpen
  \bibfield  {author} {\bibinfo {author} {\bibfnamefont {C.~A.}\ \bibnamefont {Rodríguez-Rosario}}, \bibinfo {author} {\bibfnamefont {K.}~\bibnamefont {Modi}}, \bibinfo {author} {\bibfnamefont {L.}~\bibnamefont {Mazzola}}, \ and\ \bibinfo {author} {\bibfnamefont {A.}~\bibnamefont {Aspuru-Guzik}},\ }\bibfield  {title} {\enquote {\bibinfo {title} {Unification of witnessing initial system-environment correlations and witnessing non-{M}arkovianity},}\ }\href {https://dx.doi.org/10.1209/0295-5075/99/20010} {\bibfield  {journal} {\bibinfo  {journal} {EPL}\ }\textbf {\bibinfo {volume} {99}},\ \bibinfo {pages} {20010} (\bibinfo {year} {2012})}\BibitemShut {NoStop}%
\bibitem [{\citenamefont {Luo}\ \emph {et~al.}(2012)\citenamefont {Luo}, \citenamefont {Fu},\ and\ \citenamefont {Song}}]{marko-measure-10}%
  \BibitemOpen
  \bibfield  {author} {\bibinfo {author} {\bibfnamefont {S.}~\bibnamefont {Luo}}, \bibinfo {author} {\bibfnamefont {S.}~\bibnamefont {Fu}}, \ and\ \bibinfo {author} {\bibfnamefont {H.}~\bibnamefont {Song}},\ }\bibfield  {title} {\enquote {\bibinfo {title} {Quantifying non-{M}arkovianity via correlations},}\ }\href {https://link.aps.org/doi/10.1103/PhysRevA.86.044101} {\bibfield  {journal} {\bibinfo  {journal} {Phys. Rev. A}\ }\textbf {\bibinfo {volume} {86}},\ \bibinfo {pages} {044101} (\bibinfo {year} {2012})}\BibitemShut {NoStop}%
\bibitem [{\citenamefont {Bylicka}\ \emph {et~al.}(2014)\citenamefont {Bylicka}, \citenamefont {Chruściński},\ and\ \citenamefont {Maniscalco}}]{marko-measure-11}%
  \BibitemOpen
  \bibfield  {author} {\bibinfo {author} {\bibfnamefont {B.}~\bibnamefont {Bylicka}}, \bibinfo {author} {\bibfnamefont {D.}~\bibnamefont {Chruściński}}, \ and\ \bibinfo {author} {\bibfnamefont {S.}~\bibnamefont {Maniscalco}},\ }\bibfield  {title} {\enquote {\bibinfo {title} {Non-{M}arkovianity and reservoir memory of quantum channels: a quantum information theory perspective},}\ }\href {https://doi.org/10.1038/srep05720} {\bibfield  {journal} {\bibinfo  {journal} {Sci. Rep.}\ }\textbf {\bibinfo {volume} {4}},\ \bibinfo {pages} {5720} (\bibinfo {year} {2014})}\BibitemShut {NoStop}%
\bibitem [{\citenamefont {Chru\ifmmode \acute{s}\else \'{s}\fi{}ci\ifmmode~\acute{n}\else \'{n}\fi{}ski}\ and\ \citenamefont {Maniscalco}(2014)}]{marko-measure-12}%
  \BibitemOpen
  \bibfield  {author} {\bibinfo {author} {\bibfnamefont {D.}~\bibnamefont {Chru\ifmmode \acute{s}\else \'{s}\fi{}ci\ifmmode~\acute{n}\else \'{n}\fi{}ski}}\ and\ \bibinfo {author} {\bibfnamefont {S.}~\bibnamefont {Maniscalco}},\ }\bibfield  {title} {\enquote {\bibinfo {title} {Degree of non-{M}arkovianity of quantum evolution},}\ }\href {https://link.aps.org/doi/10.1103/PhysRevLett.112.120404} {\bibfield  {journal} {\bibinfo  {journal} {Phys. Rev. Lett.}\ }\textbf {\bibinfo {volume} {112}},\ \bibinfo {pages} {120404} (\bibinfo {year} {2014})}\BibitemShut {NoStop}%
\bibitem [{\citenamefont {Bylicka}\ \emph {et~al.}(2017)\citenamefont {Bylicka}, \citenamefont {Johansson},\ and\ \citenamefont {Ac\'{\i}n}}]{marko-measure-13}%
  \BibitemOpen
  \bibfield  {author} {\bibinfo {author} {\bibfnamefont {B.}~\bibnamefont {Bylicka}}, \bibinfo {author} {\bibfnamefont {M.}~\bibnamefont {Johansson}}, \ and\ \bibinfo {author} {\bibfnamefont {A.}~\bibnamefont {Ac\'{\i}n}},\ }\bibfield  {title} {\enquote {\bibinfo {title} {Constructive method for detecting the information backflow of non-{M}arkovian dynamics},}\ }\href {https://link.aps.org/doi/10.1103/PhysRevLett.118.120501} {\bibfield  {journal} {\bibinfo  {journal} {Phys. Rev. Lett.}\ }\textbf {\bibinfo {volume} {118}},\ \bibinfo {pages} {120501} (\bibinfo {year} {2017})}\BibitemShut {NoStop}%
\bibitem [{\citenamefont {He}\ \emph {et~al.}(2017)\citenamefont {He}, \citenamefont {Zeng}, \citenamefont {Li}, \citenamefont {Wang},\ and\ \citenamefont {Yao}}]{marko-measure-14}%
  \BibitemOpen
  \bibfield  {author} {\bibinfo {author} {\bibfnamefont {Z.}~\bibnamefont {He}}, \bibinfo {author} {\bibfnamefont {H-S.}\ \bibnamefont {Zeng}}, \bibinfo {author} {\bibfnamefont {Y.}~\bibnamefont {Li}}, \bibinfo {author} {\bibfnamefont {Q.}~\bibnamefont {Wang}}, \ and\ \bibinfo {author} {\bibfnamefont {C.}~\bibnamefont {Yao}},\ }\bibfield  {title} {\enquote {\bibinfo {title} {Non-{M}arkovianity measure based on the relative entropy of coherence in an extended space},}\ }\href {https://link.aps.org/doi/10.1103/PhysRevA.96.022106} {\bibfield  {journal} {\bibinfo  {journal} {Phys. Rev. A}\ }\textbf {\bibinfo {volume} {96}},\ \bibinfo {pages} {022106} (\bibinfo {year} {2017})}\BibitemShut {NoStop}%
\bibitem [{\citenamefont {Chru\ifmmode \acute{s}\else \'{s}\fi{}ci\ifmmode~\acute{n}\else \'{n}\fi{}ski}\ \emph {et~al.}(2018)\citenamefont {Chru\ifmmode \acute{s}\else \'{s}\fi{}ci\ifmmode~\acute{n}\else \'{n}\fi{}ski}, \citenamefont {Rivas},\ and\ \citenamefont {St\o{}rmer}}]{marko-measure-15}%
  \BibitemOpen
  \bibfield  {author} {\bibinfo {author} {\bibfnamefont {D.}~\bibnamefont {Chru\ifmmode \acute{s}\else \'{s}\fi{}ci\ifmmode~\acute{n}\else \'{n}\fi{}ski}}, \bibinfo {author} {\bibfnamefont {\'A.}\ \bibnamefont {Rivas}}, \ and\ \bibinfo {author} {\bibfnamefont {E.}~\bibnamefont {St\o{}rmer}},\ }\bibfield  {title} {\enquote {\bibinfo {title} {Divisibility and information flow notions of quantum {M}arkovianity for noninvertible dynamical maps},}\ }\href {https://link.aps.org/doi/10.1103/PhysRevLett.121.080407} {\bibfield  {journal} {\bibinfo  {journal} {Phys. Rev. Lett.}\ }\textbf {\bibinfo {volume} {121}},\ \bibinfo {pages} {080407} (\bibinfo {year} {2018})}\BibitemShut {NoStop}%
\bibitem [{\citenamefont {Bennett}\ \emph {et~al.}(1996)\citenamefont {Bennett}, \citenamefont {DiVincenzo}, \citenamefont {Smolin},\ and\ \citenamefont {Wootters}}]{ent-re-1}%
  \BibitemOpen
  \bibfield  {author} {\bibinfo {author} {\bibfnamefont {C.~H.}\ \bibnamefont {Bennett}}, \bibinfo {author} {\bibfnamefont {D.~P.}\ \bibnamefont {DiVincenzo}}, \bibinfo {author} {\bibfnamefont {J.~A.}\ \bibnamefont {Smolin}}, \ and\ \bibinfo {author} {\bibfnamefont {W.~K.}\ \bibnamefont {Wootters}},\ }\bibfield  {title} {\enquote {\bibinfo {title} {Mixed-state entanglement and quantum error correction},}\ }\href {https://link.aps.org/doi/10.1103/PhysRevA.54.3824} {\bibfield  {journal} {\bibinfo  {journal} {Phys. Rev. A}\ }\textbf {\bibinfo {volume} {54}},\ \bibinfo {pages} {3824} (\bibinfo {year} {1996})}\BibitemShut {NoStop}%
\bibitem [{\citenamefont {Rains}(1998)}]{ent-re-2}%
  \BibitemOpen
  \bibfield  {author} {\bibinfo {author} {\bibfnamefont {E.~M.}\ \bibnamefont {Rains}},\ }\bibfield  {title} {\enquote {\bibinfo {title} {Entanglement purification via separable superoperators},}\ }\href {https://doi.org/10.48550/arXiv.quant-ph/9707002} {\bibfield  {journal} {\bibinfo  {journal} {arXiv:quant-ph/9707002}\ } (\bibinfo {year} {1998})}\BibitemShut {NoStop}%
\bibitem [{\citenamefont {Vedral}\ and\ \citenamefont {Plenio}(1998)}]{ent-re-3}%
  \BibitemOpen
  \bibfield  {author} {\bibinfo {author} {\bibfnamefont {V.}~\bibnamefont {Vedral}}\ and\ \bibinfo {author} {\bibfnamefont {M.~B.}\ \bibnamefont {Plenio}},\ }\bibfield  {title} {\enquote {\bibinfo {title} {Entanglement measures and purification procedures},}\ }\href {https://link.aps.org/doi/10.1103/PhysRevA.57.1619} {\bibfield  {journal} {\bibinfo  {journal} {Phys. Rev. A}\ }\textbf {\bibinfo {volume} {57}},\ \bibinfo {pages} {1619} (\bibinfo {year} {1998})}\BibitemShut {NoStop}%
\bibitem [{\citenamefont {Rains}(1999)}]{ent-re-4}%
  \BibitemOpen
  \bibfield  {author} {\bibinfo {author} {\bibfnamefont {E.~M.}\ \bibnamefont {Rains}},\ }\bibfield  {title} {\enquote {\bibinfo {title} {Bound on distillable entanglement},}\ }\href {https://link.aps.org/doi/10.1103/PhysRevA.60.179} {\bibfield  {journal} {\bibinfo  {journal} {Phys. Rev. A}\ }\textbf {\bibinfo {volume} {60}},\ \bibinfo {pages} {179} (\bibinfo {year} {1999})}\BibitemShut {NoStop}%
\bibitem [{\citenamefont {Horodecki}\ \emph {et~al.}(2009)\citenamefont {Horodecki}, \citenamefont {Horodecki}, \citenamefont {Horodecki},\ and\ \citenamefont {Horodecki}}]{ent-re-5}%
  \BibitemOpen
  \bibfield  {author} {\bibinfo {author} {\bibfnamefont {R.}~\bibnamefont {Horodecki}}, \bibinfo {author} {\bibfnamefont {P.}~\bibnamefont {Horodecki}}, \bibinfo {author} {\bibfnamefont {M.}~\bibnamefont {Horodecki}}, \ and\ \bibinfo {author} {\bibfnamefont {K.}~\bibnamefont {Horodecki}},\ }\bibfield  {title} {\enquote {\bibinfo {title} {Quantum entanglement},}\ }\href {https://link.aps.org/doi/10.1103/RevModPhys.81.865} {\bibfield  {journal} {\bibinfo  {journal} {Rev. Mod. Phys.}\ }\textbf {\bibinfo {volume} {81}},\ \bibinfo {pages} {865} (\bibinfo {year} {2009})}\BibitemShut {NoStop}%
\bibitem [{\citenamefont {Horodecki}\ \emph {et~al.}(2003)\citenamefont {Horodecki}, \citenamefont {Horodecki},\ and\ \citenamefont {Oppenheim}}]{thermo-re-2}%
  \BibitemOpen
  \bibfield  {author} {\bibinfo {author} {\bibfnamefont {M.}~\bibnamefont {Horodecki}}, \bibinfo {author} {\bibfnamefont {P.}~\bibnamefont {Horodecki}}, \ and\ \bibinfo {author} {\bibfnamefont {J.}~\bibnamefont {Oppenheim}},\ }\bibfield  {title} {\enquote {\bibinfo {title} {Reversible transformations from pure to mixed states and the unique measure of information},}\ }\href {https://link.aps.org/doi/10.1103/PhysRevA.67.062104} {\bibfield  {journal} {\bibinfo  {journal} {Phys. Rev. A}\ }\textbf {\bibinfo {volume} {67}},\ \bibinfo {pages} {062104} (\bibinfo {year} {2003})}\BibitemShut {NoStop}%
\bibitem [{\citenamefont {Gour}\ \emph {et~al.}(2015)\citenamefont {Gour}, \citenamefont {Müller}, \citenamefont {Narasimhachar}, \citenamefont {Spekkens},\ and\ \citenamefont {Halpern}}]{purity-re-2}%
  \BibitemOpen
  \bibfield  {author} {\bibinfo {author} {\bibfnamefont {G.}~\bibnamefont {Gour}}, \bibinfo {author} {\bibfnamefont {M.~P.}\ \bibnamefont {Müller}}, \bibinfo {author} {\bibfnamefont {V.}~\bibnamefont {Narasimhachar}}, \bibinfo {author} {\bibfnamefont {R.~W.}\ \bibnamefont {Spekkens}}, \ and\ \bibinfo {author} {\bibfnamefont {N.~Y.}\ \bibnamefont {Halpern}},\ }\bibfield  {title} {\enquote {\bibinfo {title} {The resource theory of informational nonequilibrium in thermodynamics},}\ }\href {https://www.sciencedirect.com/science/article/pii/S037015731500229X} {\bibfield  {journal} {\bibinfo  {journal} {Phys. Rep.}\ }\textbf {\bibinfo {volume} {583}},\ \bibinfo {pages} {1} (\bibinfo {year} {2015})}\BibitemShut {NoStop}%
\bibitem [{\citenamefont {Aberg}(2006)}]{coherence-re-1}%
  \BibitemOpen
  \bibfield  {author} {\bibinfo {author} {\bibfnamefont {J.}~\bibnamefont {Aberg}},\ }\bibfield  {title} {\enquote {\bibinfo {title} {Quantifying superposition},}\ }\href {https://doi.org/10.48550/arXiv.quant-ph/0612146} {\bibfield  {journal} {\bibinfo  {journal} {arXiv:quant-ph/0612146}\ } (\bibinfo {year} {2006})}\BibitemShut {NoStop}%
\bibitem [{\citenamefont {Baumgratz}\ \emph {et~al.}(2014)\citenamefont {Baumgratz}, \citenamefont {Cramer},\ and\ \citenamefont {Plenio}}]{coherence-re-2}%
  \BibitemOpen
  \bibfield  {author} {\bibinfo {author} {\bibfnamefont {T.}~\bibnamefont {Baumgratz}}, \bibinfo {author} {\bibfnamefont {M.}~\bibnamefont {Cramer}}, \ and\ \bibinfo {author} {\bibfnamefont {M.~B.}\ \bibnamefont {Plenio}},\ }\bibfield  {title} {\enquote {\bibinfo {title} {Quantifying coherence},}\ }\href {https://link.aps.org/doi/10.1103/PhysRevLett.113.140401} {\bibfield  {journal} {\bibinfo  {journal} {Phys. Rev. Lett.}\ }\textbf {\bibinfo {volume} {113}},\ \bibinfo {pages} {140401} (\bibinfo {year} {2014})}\BibitemShut {NoStop}%
\bibitem [{\citenamefont {Winter}\ and\ \citenamefont {Yang}(2016)}]{coherence-re-3}%
  \BibitemOpen
  \bibfield  {author} {\bibinfo {author} {\bibfnamefont {A.}~\bibnamefont {Winter}}\ and\ \bibinfo {author} {\bibfnamefont {D.}~\bibnamefont {Yang}},\ }\bibfield  {title} {\enquote {\bibinfo {title} {Operational resource theory of coherence},}\ }\href {https://link.aps.org/doi/10.1103/PhysRevLett.116.120404} {\bibfield  {journal} {\bibinfo  {journal} {Phys. Rev. Lett.}\ }\textbf {\bibinfo {volume} {116}},\ \bibinfo {pages} {120404} (\bibinfo {year} {2016})}\BibitemShut {NoStop}%
\bibitem [{\citenamefont {Gour}\ and\ \citenamefont {Spekkens}(2008)}]{asymmetry-re-1}%
  \BibitemOpen
  \bibfield  {author} {\bibinfo {author} {\bibfnamefont {G.}~\bibnamefont {Gour}}\ and\ \bibinfo {author} {\bibfnamefont {R.~W.}\ \bibnamefont {Spekkens}},\ }\bibfield  {title} {\enquote {\bibinfo {title} {The resource theory of quantum reference frames: manipulations and monotones},}\ }\href {https://dx.doi.org/10.1088/1367-2630/10/3/033023} {\bibfield  {journal} {\bibinfo  {journal} {New J. Phys.}\ }\textbf {\bibinfo {volume} {10}},\ \bibinfo {pages} {033023} (\bibinfo {year} {2008})}\BibitemShut {NoStop}%
\bibitem [{\citenamefont {Gour}\ \emph {et~al.}(2009)\citenamefont {Gour}, \citenamefont {Marvian},\ and\ \citenamefont {Spekkens}}]{asymmetry-re-2}%
  \BibitemOpen
  \bibfield  {author} {\bibinfo {author} {\bibfnamefont {G.}~\bibnamefont {Gour}}, \bibinfo {author} {\bibfnamefont {I.}~\bibnamefont {Marvian}}, \ and\ \bibinfo {author} {\bibfnamefont {R.~W.}\ \bibnamefont {Spekkens}},\ }\bibfield  {title} {\enquote {\bibinfo {title} {Measuring the quality of a quantum reference frame: The relative entropy of frameness},}\ }\href {https://link.aps.org/doi/10.1103/PhysRevA.80.012307} {\bibfield  {journal} {\bibinfo  {journal} {Phys. Rev. A}\ }\textbf {\bibinfo {volume} {80}},\ \bibinfo {pages} {012307} (\bibinfo {year} {2009})}\BibitemShut {NoStop}%
\bibitem [{\citenamefont {Marvian}\ and\ \citenamefont {Spekkens}(2013)}]{asymmetry-re-3}%
  \BibitemOpen
  \bibfield  {author} {\bibinfo {author} {\bibfnamefont {I.}~\bibnamefont {Marvian}}\ and\ \bibinfo {author} {\bibfnamefont {R.~W.}\ \bibnamefont {Spekkens}},\ }\bibfield  {title} {\enquote {\bibinfo {title} {The theory of manipulations of pure state asymmetry: I. basic tools, equivalence classes and single copy transformations},}\ }\href {https://dx.doi.org/10.1088/1367-2630/15/3/033001} {\bibfield  {journal} {\bibinfo  {journal} {New J. Phys.}\ }\textbf {\bibinfo {volume} {15}},\ \bibinfo {pages} {033001} (\bibinfo {year} {2013})}\BibitemShut {NoStop}%
\bibitem [{\citenamefont {Mari}\ and\ \citenamefont {Eisert}(2012)}]{magic-re-1}%
  \BibitemOpen
  \bibfield  {author} {\bibinfo {author} {\bibfnamefont {A.}~\bibnamefont {Mari}}\ and\ \bibinfo {author} {\bibfnamefont {J.}~\bibnamefont {Eisert}},\ }\bibfield  {title} {\enquote {\bibinfo {title} {Positive wigner functions render classical simulation of quantum computation efficient},}\ }\href {https://link.aps.org/doi/10.1103/PhysRevLett.109.230503} {\bibfield  {journal} {\bibinfo  {journal} {Phys. Rev. Lett.}\ }\textbf {\bibinfo {volume} {109}},\ \bibinfo {pages} {230503} (\bibinfo {year} {2012})}\BibitemShut {NoStop}%
\bibitem [{\citenamefont {Veitch}\ \emph {et~al.}(2012)\citenamefont {Veitch}, \citenamefont {Ferrie}, \citenamefont {Gross},\ and\ \citenamefont {Emerson}}]{magic-re-2}%
  \BibitemOpen
  \bibfield  {author} {\bibinfo {author} {\bibfnamefont {V.}~\bibnamefont {Veitch}}, \bibinfo {author} {\bibfnamefont {C.}~\bibnamefont {Ferrie}}, \bibinfo {author} {\bibfnamefont {D.}~\bibnamefont {Gross}}, \ and\ \bibinfo {author} {\bibfnamefont {J.}~\bibnamefont {Emerson}},\ }\bibfield  {title} {\enquote {\bibinfo {title} {Negative quasi-probability as a resource for quantum computation},}\ }\href {https://dx.doi.org/10.1088/1367-2630/14/11/113011} {\bibfield  {journal} {\bibinfo  {journal} {New J. Phys.}\ }\textbf {\bibinfo {volume} {14}},\ \bibinfo {pages} {113011} (\bibinfo {year} {2012})}\BibitemShut {NoStop}%
\bibitem [{\citenamefont {Veitch}\ \emph {et~al.}(2014)\citenamefont {Veitch}, \citenamefont {Mousavian}, \citenamefont {Gottesman},\ and\ \citenamefont {Emerson}}]{magic-re-3}%
  \BibitemOpen
  \bibfield  {author} {\bibinfo {author} {\bibfnamefont {V.}~\bibnamefont {Veitch}}, \bibinfo {author} {\bibfnamefont {S.~A.~H.}\ \bibnamefont {Mousavian}}, \bibinfo {author} {\bibfnamefont {D.}~\bibnamefont {Gottesman}}, \ and\ \bibinfo {author} {\bibfnamefont {J.}~\bibnamefont {Emerson}},\ }\bibfield  {title} {\enquote {\bibinfo {title} {The resource theory of stabilizer quantum computation},}\ }\href {https://dx.doi.org/10.1088/1367-2630/16/1/013009} {\bibfield  {journal} {\bibinfo  {journal} {New J. Phys.}\ }\textbf {\bibinfo {volume} {16}},\ \bibinfo {pages} {013009} (\bibinfo {year} {2014})}\BibitemShut {NoStop}%
\bibitem [{\citenamefont {Janzing}\ \emph {et~al.}(2000)\citenamefont {Janzing}, \citenamefont {Wocjan}, \citenamefont {Zeier}, \citenamefont {Geiss},\ and\ \citenamefont {Beth}}]{thermo-re-1}%
  \BibitemOpen
  \bibfield  {author} {\bibinfo {author} {\bibfnamefont {D.}~\bibnamefont {Janzing}}, \bibinfo {author} {\bibfnamefont {P.}~\bibnamefont {Wocjan}}, \bibinfo {author} {\bibfnamefont {R.}~\bibnamefont {Zeier}}, \bibinfo {author} {\bibfnamefont {R.}~\bibnamefont {Geiss}}, \ and\ \bibinfo {author} {\bibfnamefont {Th.}\ \bibnamefont {Beth}},\ }\bibfield  {title} {\enquote {\bibinfo {title} {Thermodynamic cost of reliability and low temperatures: Tightening landauer's principle and the second law},}\ }\href {https://doi.org/10.1023/A:1026422630734} {\bibfield  {journal} {\bibinfo  {journal} {Int. J. Theor. Phys.}\ }\textbf {\bibinfo {volume} {39}},\ \bibinfo {pages} {2717} (\bibinfo {year} {2000})}\BibitemShut {NoStop}%
\bibitem [{\citenamefont {Rio}\ \emph {et~al.}(2011)\citenamefont {Rio}, \citenamefont {Åberg}, \citenamefont {Renner}, \citenamefont {Dahlsten},\ and\ \citenamefont {Vedral}}]{thermo-re-3}%
  \BibitemOpen
  \bibfield  {author} {\bibinfo {author} {\bibfnamefont {L.~del}\ \bibnamefont {Rio}}, \bibinfo {author} {\bibfnamefont {J.}~\bibnamefont {Åberg}}, \bibinfo {author} {\bibfnamefont {R.}~\bibnamefont {Renner}}, \bibinfo {author} {\bibfnamefont {O.}~\bibnamefont {Dahlsten}}, \ and\ \bibinfo {author} {\bibfnamefont {V.}~\bibnamefont {Vedral}},\ }\bibfield  {title} {\enquote {\bibinfo {title} {The thermodynamic meaning of negative entropy},}\ }\href {https://doi.org/10.1038/nature10123} {\bibfield  {journal} {\bibinfo  {journal} {Nature}\ }\textbf {\bibinfo {volume} {474}},\ \bibinfo {pages} {61} (\bibinfo {year} {2011})}\BibitemShut {NoStop}%
\bibitem [{\citenamefont {Dahlsten}\ \emph {et~al.}(2011)\citenamefont {Dahlsten}, \citenamefont {Renner}, \citenamefont {Rieper},\ and\ \citenamefont {Vedral}}]{thermo-re-4}%
  \BibitemOpen
  \bibfield  {author} {\bibinfo {author} {\bibfnamefont {O.~C.~O.}\ \bibnamefont {Dahlsten}}, \bibinfo {author} {\bibfnamefont {R.}~\bibnamefont {Renner}}, \bibinfo {author} {\bibfnamefont {E.}~\bibnamefont {Rieper}}, \ and\ \bibinfo {author} {\bibfnamefont {V.}~\bibnamefont {Vedral}},\ }\bibfield  {title} {\enquote {\bibinfo {title} {Inadequacy of von neumann entropy for characterizing extractable work},}\ }\href {https://dx.doi.org/10.1088/1367-2630/13/5/053015} {\bibfield  {journal} {\bibinfo  {journal} {New J. Phys.}\ }\textbf {\bibinfo {volume} {13}},\ \bibinfo {pages} {053015} (\bibinfo {year} {2011})}\BibitemShut {NoStop}%
\bibitem [{\citenamefont {Brand\~ao}\ \emph {et~al.}(2013)\citenamefont {Brand\~ao}, \citenamefont {Horodecki}, \citenamefont {Oppenheim}, \citenamefont {Renes},\ and\ \citenamefont {Spekkens}}]{thermo-re-5}%
  \BibitemOpen
  \bibfield  {author} {\bibinfo {author} {\bibfnamefont {F.~G. S.~L.}\ \bibnamefont {Brand\~ao}}, \bibinfo {author} {\bibfnamefont {M.}~\bibnamefont {Horodecki}}, \bibinfo {author} {\bibfnamefont {J.}~\bibnamefont {Oppenheim}}, \bibinfo {author} {\bibfnamefont {J.~M.}\ \bibnamefont {Renes}}, \ and\ \bibinfo {author} {\bibfnamefont {R.~W.}\ \bibnamefont {Spekkens}},\ }\bibfield  {title} {\enquote {\bibinfo {title} {Resource theory of quantum states out of thermal equilibrium},}\ }\href {https://link.aps.org/doi/10.1103/PhysRevLett.111.250404} {\bibfield  {journal} {\bibinfo  {journal} {Phys. Rev. Lett.}\ }\textbf {\bibinfo {volume} {111}},\ \bibinfo {pages} {250404} (\bibinfo {year} {2013})}\BibitemShut {NoStop}%
\bibitem [{\citenamefont {Horodecki}\ and\ \citenamefont {Oppenheim}(2013)}]{thermo-re-6}%
  \BibitemOpen
  \bibfield  {author} {\bibinfo {author} {\bibfnamefont {M.}~\bibnamefont {Horodecki}}\ and\ \bibinfo {author} {\bibfnamefont {J.}~\bibnamefont {Oppenheim}},\ }\bibfield  {title} {\enquote {\bibinfo {title} {Fundamental limitations for quantum and nanoscale thermodynamics},}\ }\href {https://doi.org/10.1038/ncomms3059} {\bibfield  {journal} {\bibinfo  {journal} {Nat. Commun}\ }\textbf {\bibinfo {volume} {4}},\ \bibinfo {pages} {2059} (\bibinfo {year} {2013})}\BibitemShut {NoStop}%
\bibitem [{\citenamefont {Skrzypczyk}\ \emph {et~al.}(2014)\citenamefont {Skrzypczyk}, \citenamefont {Short},\ and\ \citenamefont {Popescu}}]{thermo-re-7}%
  \BibitemOpen
  \bibfield  {author} {\bibinfo {author} {\bibfnamefont {P.}~\bibnamefont {Skrzypczyk}}, \bibinfo {author} {\bibfnamefont {A.~J.}\ \bibnamefont {Short}}, \ and\ \bibinfo {author} {\bibfnamefont {S.}~\bibnamefont {Popescu}},\ }\bibfield  {title} {\enquote {\bibinfo {title} {Work extraction and thermodynamics for individual quantum systems},}\ }\href {https://doi.org/10.1038/ncomms5185} {\bibfield  {journal} {\bibinfo  {journal} {Nat. Commun}\ }\textbf {\bibinfo {volume} {5}},\ \bibinfo {pages} {4185} (\bibinfo {year} {2014})}\BibitemShut {NoStop}%
\bibitem [{\citenamefont {Gallego}\ \emph {et~al.}(2016)\citenamefont {Gallego}, \citenamefont {Eisert},\ and\ \citenamefont {Wilming}}]{thermo-re-8}%
  \BibitemOpen
  \bibfield  {author} {\bibinfo {author} {\bibfnamefont {R.}~\bibnamefont {Gallego}}, \bibinfo {author} {\bibfnamefont {J.}~\bibnamefont {Eisert}}, \ and\ \bibinfo {author} {\bibfnamefont {H.}~\bibnamefont {Wilming}},\ }\bibfield  {title} {\enquote {\bibinfo {title} {Thermodynamic work from operational principles},}\ }\href {https://dx.doi.org/10.1088/1367-2630/18/10/103017} {\bibfield  {journal} {\bibinfo  {journal} {New J. Phys.}\ }\textbf {\bibinfo {volume} {18}},\ \bibinfo {pages} {103017} (\bibinfo {year} {2016})}\BibitemShut {NoStop}%
\bibitem [{\citenamefont {Chitambar}\ and\ \citenamefont {Gour}(2019)}]{re-1}%
  \BibitemOpen
  \bibfield  {author} {\bibinfo {author} {\bibfnamefont {E.}~\bibnamefont {Chitambar}}\ and\ \bibinfo {author} {\bibfnamefont {G.}~\bibnamefont {Gour}},\ }\bibfield  {title} {\enquote {\bibinfo {title} {Quantum resource theories},}\ }\href {https://link.aps.org/doi/10.1103/RevModPhys.91.025001} {\bibfield  {journal} {\bibinfo  {journal} {Rev. Mod. Phys.}\ }\textbf {\bibinfo {volume} {91}},\ \bibinfo {pages} {025001} (\bibinfo {year} {2019})}\BibitemShut {NoStop}%
\bibitem [{\citenamefont {Gour}(2024)}]{re-book-2}%
  \BibitemOpen
  \bibfield  {author} {\bibinfo {author} {\bibfnamefont {G.}~\bibnamefont {Gour}},\ }\bibfield  {title} {\enquote {\bibinfo {title} {Resources of the quantum world},}\ }\href {https://doi.org/10.48550/arXiv.2402.05474} {\bibfield  {journal} {\bibinfo  {journal} {arXiv:2402.05474}\ } (\bibinfo {year} {2024})}\BibitemShut {NoStop}%
\bibitem [{\citenamefont {Chru\ifmmode \acute{s}\else \'{s}\fi{}ci\ifmmode~\acute{n}\else \'{n}\fi{}ski}\ \emph {et~al.}(2011)\citenamefont {Chru\ifmmode \acute{s}\else \'{s}\fi{}ci\ifmmode~\acute{n}\else \'{n}\fi{}ski}, \citenamefont {Kossakowski},\ and\ \citenamefont {Rivas}}]{P-div-inf-backflow}%
  \BibitemOpen
  \bibfield  {author} {\bibinfo {author} {\bibfnamefont {D.}~\bibnamefont {Chru\ifmmode \acute{s}\else \'{s}\fi{}ci\ifmmode~\acute{n}\else \'{n}\fi{}ski}}, \bibinfo {author} {\bibfnamefont {A.}~\bibnamefont {Kossakowski}}, \ and\ \bibinfo {author} {\bibfnamefont {\'A.}\ \bibnamefont {Rivas}},\ }\bibfield  {title} {\enquote {\bibinfo {title} {Measures of non-{M}arkovianity: Divisibility versus backflow of information},}\ }\href {https://link.aps.org/doi/10.1103/PhysRevA.83.052128} {\bibfield  {journal} {\bibinfo  {journal} {Phys. Rev. A}\ }\textbf {\bibinfo {volume} {83}},\ \bibinfo {pages} {052128} (\bibinfo {year} {2011})}\BibitemShut {NoStop}%
\bibitem [{\citenamefont {Lidar}(2020)}]{lidar-notes}%
  \BibitemOpen
  \bibfield  {author} {\bibinfo {author} {\bibfnamefont {D.~A.}\ \bibnamefont {Lidar}},\ }\bibfield  {title} {\enquote {\bibinfo {title} {Lecture notes on the theory of open quantum systems},}\ }\href {https://doi.org/10.48550/arXiv.1902.00967} {\bibfield  {journal} {\bibinfo  {journal} {arXiv:1902.00967}\ } (\bibinfo {year} {2020})}\BibitemShut {NoStop}%
\end{thebibliography}%

\end{document}